\documentclass[useAMS,usenatbib]{mn2e}
\usepackage{graphicx}
\usepackage{txfonts} 
\usepackage{amsmath}

\newcommand{\aap}{A\&A}
\newcommand{\mnras}{MNRAS}
\newcommand{\apj}{ApJ}
\newcommand{\aj}{AJ}

\newcommand{\apjs}{ApJS}
\newcommand{\araa}{ARA\&A}
\newcommand{\nat}{Nature}

\catcode`\@=11
\def\gsim{\ifmmode{\mathrel{\mathpalette\@versim>}}
    \else{$\mathrel{\mathpalette\@versim>}$}\fi}
\def\lsim{\ifmmode{\mathrel{\mathpalette\@versim<}}
    \else{$\mathrel{\mathpalette\@versim<}$}\fi}
\def\@versim#1#2{\lower 2.9truept \vbox{\baselineskip 0pt \lineskip
    0.5truept \ialign{$\m@th#1\hfil##\hfil$\crcr#2\crcr\sim\crcr}}}
\catcode`\@=12
\def\msun{\hbox{$M_\odot$}}

\def\yr-1{\hbox{${\rm yr}^{-1}$}}

\def\zf{\hbox{$z_{\rm F}$}}

\def\t9{\hbox{$t_9$}}

\def\m*{\hbox{$M_{\rm stars}$}}

\def\ho{\hbox{$H_\circ$}}
\def\h50{\hbox{$\ho /50$}}

\def\zf{\hbox{$z_{\rm f}$}}
\def\tf{\hbox{$t_{\rm f}$}}
\def\zmin{\hbox{$z_{\rm min}$}}
\begin{document}

\title{Search Instructions for  Globular Clusters in Formation at High Redshifts }
\author[Lucia Pozzetti, Claudia Maraston  and Alvio Renzini]{Lucia Pozzetti$^{1}$\thanks{E-mail: lucia.pozzetti@inaf.it}, Claudia Maraston$^{2}$\thanks{E-mail:  Claudia.Maraston@port.ac.uk} and Alvio Renzini$^{3}$\thanks{E-mail: 
alvio.renzini@inaf.it}\\ 
$^{1}$INAF - Osservatorio di Astrofisica e Scienza dello Spazio di Bologna
 via Gobetti 93/3, I-40129 Bologna, Italy \\
$^{2}$Institute of Cosmology and Gravitation, University of Portsmouth, Burnaby Road, Portsmouth, PO1 3FX, UK\\
 $^{3}$INAF - Osservatorio
Astronomico di Padova, Vicolo dell'Osservatorio 5, I-35122 Padova,
Italy}

\date{Accepted ... 2018; Received July, 2018in original form}
 \pagerange{\pageref{firstpage}--\pageref{lastpage}} \pubyear{2002}

\maketitle
                                                            
\label{firstpage}

\begin{abstract}
The formation of globular clusters (GC), with their multiple stellar generations, is still an unsolved puzzle. Thus, interest is rising on the possibility  to detect their precursors at high redshift, hence directly witnessing their formation. A simple set of assumptions are empirically justified and then used to predict how many such precursors  formed between redshift 3 and 10 could actually be detected by the NIRCam instrument on board of JWST.  It is shown that the near power-law shape of the rest-frame UV continuum of young globular cluster precursors (GCP) implies that both colours and luminosities in NIRCam long-wavelength passbands  depend remarkably weakly on formation redshift. Thus, the predicted number counts depend only little on the actual 
formation redshifts in the mentioned range, with the exception of the bluest passbands for which counts can be  strongly suppressed by intergalactic absorption along the line of sight. Instead, counts depend strongly 
on the actual mass of GCPs, in such a way that one NIRCam pointing should detect of order of 10 GCPs to mag$\sim 30$ if their mass distribution was the same of today 
GCs, or over 1,000 if their mass was 10 times higher. Therefore, GCP number counts will set fairly tight constraints on the initial mass of GCs.
An encouraging agreement with the number density of candidate GCPs at $z=6-8$, revealed by the Hubble Frontier Fields (HFF) program, suggests that their initial mass could be at least 4 times higher than that of their local descendants if all were to end up as GCs.
\end{abstract}


\maketitle

\section{Introduction}
\label{intro}
The formation of globular clusters (GC) along with their multiple stellar populations remains a major unsolved issue in astrophysics. A new opportunity to attack the problem has recently emerged in view of the James Webb Space Telescope (JWST) operations, i.e., the direct observation of forming GCs at high redshifts. Actually, the first suggestion of its  possible feasibility is quite old, with \cite{carlberg02} having made early predictions on the expected luminosity function and clustering of high-redshift GCs (up to $z\lsim 10$). More recently, the observability of GCs in formation at high redshift has been addressed by \cite{katz13}, \cite{trenti15}, \cite{renzini17} and \cite{zick18}, whereas there are hints that some GC precursors (GCP) may have been already detected (\citealt{vanzella16,vanzella17a,vanzella17b,bouwens18}). 
As emphasised in these papers, the search for  {\it first galaxies}, GCPs and the agents of cosmic reionization are tightly  interconnected from an observational point of view, and also include  a possible direct role of GCPs in the reionisation (see also \citealt{ricotti02,schraerer11,boylan18}).

In this paper we present some of the expected properties of GCPs at high redshifts, such as luminosities, colours and luminosity functions, specifically for the passbands of the Near Infrared Camera (NIRCam)  on board of JWST.  In doing so we capitalise on the most salient properties of GCs in our Galaxy and in other galaxies in the local Universe, including their old age, broad metallicity distribution, mass function, compactness and puzzling multiple populations.
These properties are succinctly summarised here.

The bulk of GCs are assigned ages of 12.5$\pm 1$ Gyr, with a possible trend of metal rich ones being slightly ($\sim 1$ Gyr) younger (\citealt{marin09, vandenberg13, brown14}).
A few clusters some Gyrs younger than this also exist, though they tend to have masses lower than typical GCs \citep{marin09}. Here we ignore this minor  component and consider the mentioned age range as encompassing virtually all GCs in the Milky Way (MW). We also assume that this age range applies to the bulk of GCs, not only in the MW but also in the local Universe as a whole (see e.g., \citealt{puzia05}). A lookback time of 12.5 Gyr corresponds to $z\simeq 5$ and when accounting for the $\pm 1$ Gyr uncertainty the redshift range we consider for the GC formation epoch becomes $3\lsim z\lsim 10$. It has been argued, and there is some evidence in support, that massive star clusters form in 
gas-rich mergers even today  \citep{zepf92}, though the old age of GCs in the MW suggests that our Galaxy did not experience much merger-driven GC formation  in the last 10 Gyr, or more.
We also note that about 20\% Galactic GCs have ages younger than $\sim 11$ Gyr according to \cite{marin09} and \cite{vandenberg13}, hence their progenitors would be found at redshifts lower than $\sim 3$, near the peak of cosmic star formation density at $z\sim 2$ \citep{madau14}. 
GCs forming at this epoch ($z\sim 2$) were likely embedded in a metal-rich, high-extinction  environment, hence more difficult to detect. 
Moreover, many of these younger GCs are very sparsely populated, hence are irrelevant in the present contest, unless they were orders of magnitude more massive at formation. 
For these reasons, we focus on the redshift range beyond 3, as that offering the best chances for GCP detection.

The present mass function of GCs, both in the MW and  in other studied galaxies, is well represented by a lognormal distribution.
From \cite{harris13}  we adopt 1.5 Mpc$^{-3}$ for the local number density of GCs, with their mass distribution peaking at $\sim 2\times 10^5$ M$_\odot$. For the distribution itself, we adopt the Gaussian distribution in log($M$) as from \cite{harris14}, their Eq. (1), and use  it to describe the mass distribution:

\begin {equation}
{dN\over dlogM}=N_\circ\, {\rm exp}[-{({\rm log}M-{\rm log}M^*)^2\over 2\times 0.52^2}]
\end {equation}

\noindent where $M^*$ is the mass at the peak of the Gaussian,  with ${\rm log}M^*/\msun =5.3$ and 0.52 is the $\sigma$ of the Gaussian,  as from Fig. 4 in \cite{harris14}.
Integrating this distribution from $-\infty$ to $+\infty$ and setting it to 1.5 Mpc$^{-3}$ one gets the normalisation $N_\circ = 1.15$  Mpc$^{-3}$.
Adopting a GCP mass function with the same Gaussian shape as that of local GCs is a conservative assumption, in principle giving a lower limit to the expected number counts of GCPs. Indeed, it has been argued that the mass function at formation may have been much different from that of the GCs surviving today, such as a power law (e.g., \citealt{fall77,fall01,gnedin97,vesperini98}). However, disruption is expected to affect predominantly lower mass clusters, hence fainter GCPs that may well be below detectability even with JWST.  In practice, the precise shape of the CGP mass function below the peak is completely irrelevant in the present context.

For our reference case, we assume that the mass function of GCPs has exactly this shape, however with the peak mass $M^*$ being 10 times higher than in the local Universe, i.e., $\sim 2\times 10^6\,\msun$. This hypothetically higher value of $M^*$ at GC formation is meant to comply with  the widely invoked necessity of GCs being substantially more massive at birth in order to account for the multiple populations that are ubiquitous among MW GCs (more below). One generally refers to it as the ``mass budget problem"  (e.g., \citealt{renzini15}, and references therein).
We emphasize that we are not arguing for the mass budget factor to be 10, as this value is used only for illustrative purposes,
with the understanding that its actual value can only be established by future observations, in particular with NIRCam on board JWST. Our  paper is meant to provide 
easily scalable predictions that future observations can test, hence setting direct observational constraints on the actual value of the mass budget factor.
The assumption for the factor of order of 10 upscale of the mass of GCPs can be supported by the following arguments. The mere mass loss from individual stars (stellar winds and supernovae) accounts for a $\sim 40$ percent mass reduction from formation to the present. On top of it, star losses via evaporation, tidal interactions and the like would account for a further mass reduction that is difficult to quantify and that depends on the structure of the GCPs that may have been different from that of the surviving GCs. For example, it has been argued that all GCs, or at least the metal poor ones, may have formed as nuclei of dwarf galaxies with most such hosts  having later dissolved with their bare nuclei  becoming the GCs of today \citep{searle78}. In this respect, the Fornax DSph galaxy and the Sagittarius galaxy, with their exceptionally high GC frequency (e.g., \citealt{brodie06}, see also \citealt{georgiev10}) lend some support to this notion.  Highly reminiscent of this scenario is the recent finding at $z\simeq 6$ of a star-forming dwarf with a size of $\sim 400$ pc and a stellar mass of $\sim 2\times 10^7\,\msun$ hosting a compact, unresolved nucleus with $R_{\rm e}\lsim 13$ pc  and a mass of $\sim 10^6\,\msun$, perhaps the best example so far of a GCP \citep{vanzella19}. Moreover, a substantially higher mass for GCPs has been invoked by virtually all scenarios for the formation of the multiple population phenomenon, though none of such scenarios is able to account for all the complexities of the observational evidence. 

An upper limit to the mass budget factor is set by considering Galactic GCs in the context of the Galactic stellar halo. The total mass in halo GCs is $\sim 3\times 10^7\,\msun$ and the mass of the halo is about 30 times higher. So, even if the whole halo was formed by stripped GCPs, the GCPs could not have been more than $\sim 30$ times more massive than the combined present mass of halo GCs.
In any event, the size of this adopted mass upscale is perhaps the most important unknown quantity that high redshift observations may allow us to measure, with the understanding that only GCPs in the high-mass portion of the distribution will have a chance to be detected, as we shall see in the sequel. 

In the MW the metallicity distribution of GCs is very broad, from less than 1/100 solar to nearly solar, with a hint of bimodality  that is evident in the GC families of massive external galaxies  (e.g., \citealt{brodie06,harris10}). In the MW the metal rich GCs, with, say [Fe/H]$\gsim -0.5$, are confined within the bulge (e.g., \citealt{barbuy98}) and even the most metal rich ones appear to be coeval with the  bulge itself \citep{ortolani95}. A few GCs with super-solar metallicity may well exist in other massive galaxies.
Now, the metal poor GCs must pre-date the formation of the major part of the body of today's host galaxies, which will certainly help their detection also for being virtually unobscured by dust not unlike faint very high redshift galaxies with their steep UV continuum \citep{bouwens14,bouwens15,vanzella19}. Conversely, metal rich GCPs must have formed only after a substantial galaxy was already in place, having reached high metallicities and therefore the young GCPs are likely to have been deeply embedded in dust, hence substantially extincted   in the UV, and therefore much more difficult to detect. In summary, our best chances to detect GCPs at high redshift are offered by the metal poor $\sim 50$ percent fraction of the total population and of it by those in the high-mass side of the distribution. 

All studied GCs harbour multiple stellar populations of various complexity, as most vividly illustrated by Hubble Space Telescope (HST) multiband photometry (e.g., \citealt{piotto15,milone17}). The  natural interpretation of the multiple population phenomenon is in terms of successive stellar generation, i.e., as a series of two or more burst of star formation, with second generation bursts being even stronger than the first one in most massive GCs \citep{milone17}. We will not try to model such multiple bursts, but following \cite{elmegreen17} we assume that at least the main burst is completed in less than $\sim 1$ Myr, a time shorter than the evolutionary times of massive stars 
($\sim 3$ to $\sim 30$ Myr, depending on mass). Such a short time is a direct consequence of the compactness of GCs, for which Elmegreen estimates a free fall time of 
$\sim 0.03$ Myr with star formation being completed in $\sim 0.3$ Myr. Thus, for the purposes of this paper, this argument allows us to approximate GCPs with simple stellar populations (SSP), i.e., a set of coeval, chemically homogeneous  stars, though we know that they are not. We shall return on this point in Section \ref{sec:caveats}.

In summary, the number density of GCPs we are going to estimate refer to precisely the precursors of those objects that we recognise as GCs in the local Universe and whose mass function is given by Eq. (1), as adopted from \cite{harris14}. It may well be that in the early Universe  objects existed similar to such CGPs but which have disappeared in the meantime. We are not trying to include such objects (see \citealt{carlberg02} for an attempt to do so) and therefore the estimates presented in this paper can be regarded as lower limits. Even so, the local volume density of GCs (1.5 Mpc$^{-3}$) implies that one  single frame of NIRCam will include over 200,000
GCPs in the redshift range 3 to 10 \citep{renzini17}, caught in whatever stage of their formation and evolution, from being still a gas cloud before forming stars, to be at the peak of its star formation rate, to possibly having already dimmed below detectability.  The question is, how many of them could be caught as bright enough to be detected?  

The standard cosmology ($H_0 =70$ km s$^{-1}$ Mpc$^{-1}$, $\Omega_{\rm m}=0.3$, $\Omega_\Lambda =0.7$) and AB magnitudes are adopted.

\section{The model SSP{\MakeLowercase{s}}}

We use the stellar population models of \cite{maraston05}, hereafter M05\footnote{www.maraston.eu}, to describe the early spectroscopic and photometric evolution of  the progenitors of present-day globular clusters, assuming they formed in the redshift range  $3< z<10$. Though these models are available for ages from 0 to 15 Gyr, in this work we shall focus at most on the first billion year of evolution as   older models  fade below JWST detection limits (as we shall show later). For simplicity we consider only one chemical composition, namely a fractional abundance of heavy elements $Z$~as $[Z/H]=-1.35$, which lies near the middle of the metallicity distribution of present-day Milky Way GCs \citep{harris10}. In any event, at the stellar ages of interest here the opacity in the envelope of massive young stars is dominated by electron scattering, hence the spectral energy distribution (SED) of young SSPs is fairly insensitive to metallicity.
 Finally, for all models we adopt the initial mass function (IMF) of \cite{chabrier03}, from 0.1 to 100 $\msun$.

\subsection{Rest-frame spectral and photometric evolution}
\begin{figure}
\includegraphics[width=84mm]{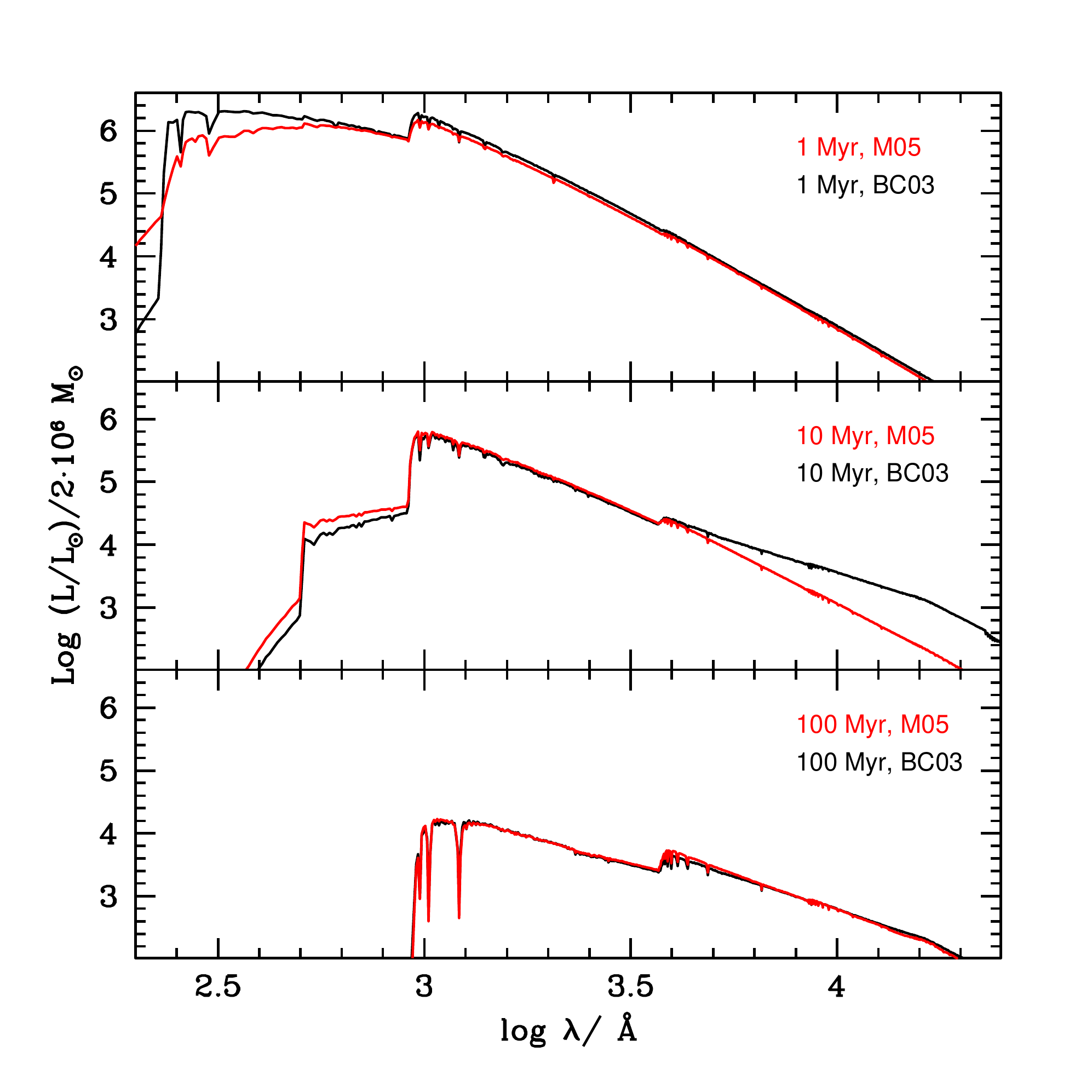}
\caption{The spectral energy distributions of SSP models for  three ages representative of very young star clusters, namely 1, 10 and 100 Myr (from top to bottom). Red and black curves refer to models by M05 and BC03, respectively, with the same IMF, and similar metallicity ($Z=10^{-3}$~and $Z=4\times10^{-3}$), respectively.}
\label{fig:sedrest}
\end{figure}

In Figure \ref{fig:sedrest} we show the rest-frame spectra of three selected models for ages that are relevant to this work (1, 10 and 100 Myr, from top to bottom) and compare them to analogue SSP models  by \cite{bruzual03}, hereafter BC03, which are based on different stellar evolutionary tracks and on the same library of stellar spectra as M05. For completeness we display the models over a wide wavelength range (90 to 25,000 \AA), but  note that differences between them at the shortest wavelengths or in the rest-frame near-IR are not relevant in the present context . Below Ly$\alpha$  the flux is absorbed by the intergalactic medium at high-$z$, while the rest-frame near-IR lies outside the NIRCam range. 
In the range $\sim 1000-9000$~\AA\ rest-frame, which is the one sampled by NIRCam passbands at the redshifts of interest, the models are very similar, hence our predictions would have been the same if using the BC03 stellar population models.

 \begin{table*}
\caption{ Example of observer-frame magnitudes in JWST filters for a GCPs of mass $log(M/M_{\odot})=6.3$. The first column gives the redshift at which the object is observed, having already aged about 3 Myr since its formation, hence formed at a slightly higher redshift. The full table can be found at: $https://sites.google.com/inaf.it/pozzetti-gcps/home$.}

\begin{tabular}{c c c c c c c c c c} \hline \\
redshift &  log(age/yr)  &  $m_{F070W}$ &  $m_{F090W}$ &  $m_{F115W}$ &  $m_{F150W}$ &  $m_{F200W}$ &  $m_{F277W}$ &  $m_{F356W}$ &  $m_{F444W}$ \\
\\
 \hline\\
  10.00 &   6.50 &  32.02 &  31.69 &  30.22 &  30.29 &  30.54 &  30.89 &  31.23 &  31.37 \\ 
   9.00 &   6.50 &  31.72 &  30.78 &  30.04 &  30.22 &  30.48 &  30.85 &  31.16 &  31.27 \\ 
   8.00 &   6.50 &  31.42 &  30.03 &  29.92 &  30.14 &  30.41 &  30.82 &  31.02 &  31.24 \\ 
   7.00 &   6.50 &  30.58 &  29.65 &  29.80 &  30.04 &  30.34 &  30.75 &  30.91 &  31.22 \\ 
   6.00 &   6.50 &  29.57 &  29.47 &  29.67 &  29.93 &  30.27 &  30.58 &  30.86 &  31.20 \\ 
   5.00 &   6.50 &  29.12 &  29.30 &  29.53 &  29.81 &  30.18 &  30.41 &  30.81 &  31.15 \\ 
   4.00 &   6.50 &  28.86 &  29.09 &  29.34 &  29.67 &  29.91 &  30.31 &  30.73 &  31.05 \\ 
   3.00 &   6.50 &  28.57 &  28.81 &  29.11 &  29.38 &  29.65 &  30.17 &  30.57 &  30.89 \\ 
   2.00 &   6.50 &  28.10 &  28.41 &  28.63 &  28.90 &  29.35 &  29.86 &  30.28 &  30.65 \\ 
\hline
\end{tabular}
\label{tab:tabellone}
\end{table*}

 \begin{figure}
\includegraphics[width=84mm]{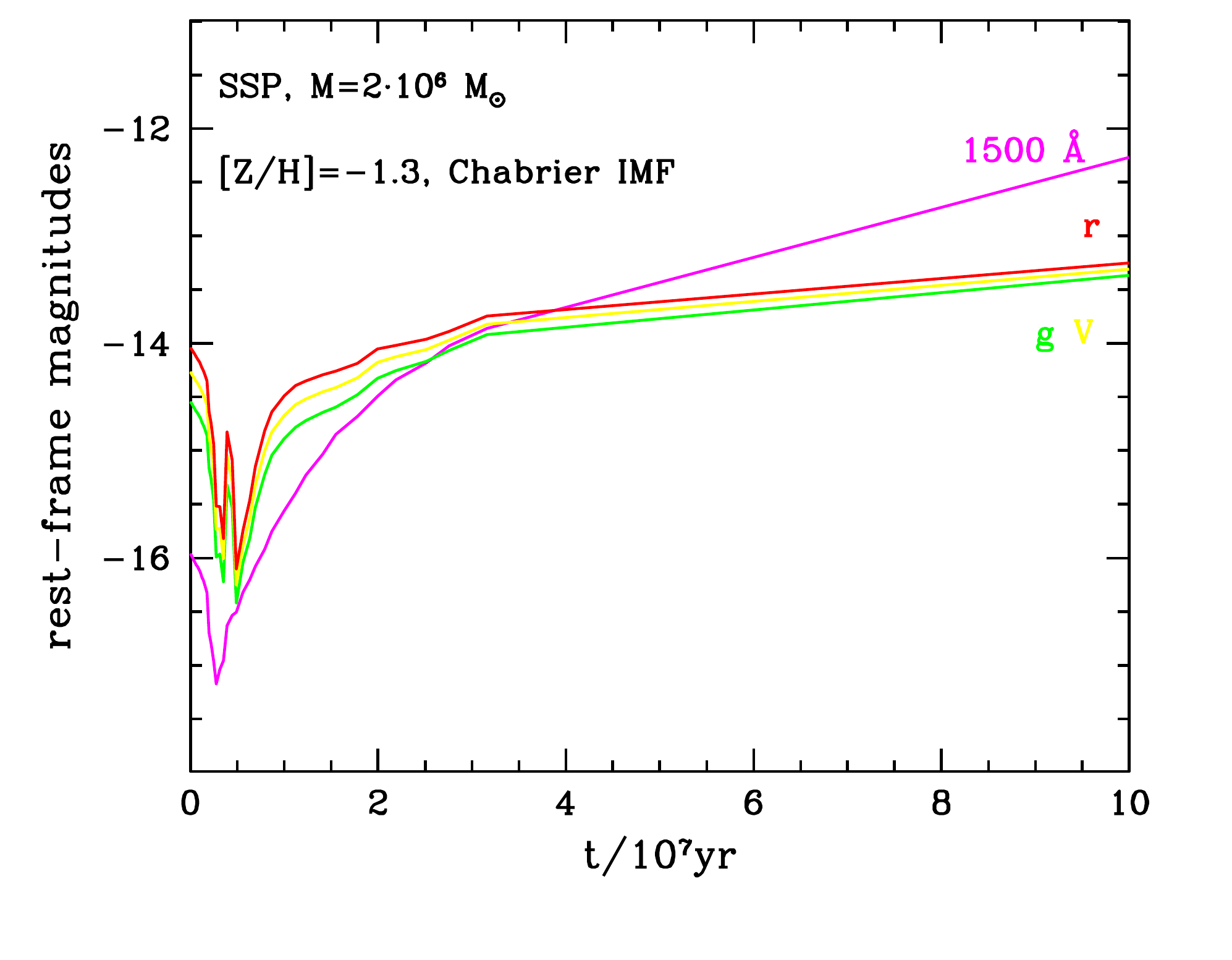}
\caption{Rest-frame absolute AB magnitudes at $1500$ \AA\ and for the $r,\,g$~and $V$~filter passbands for the SSP models considered here, shown for $t<100$~Myr and our typical total stellar mass $M^*$ of $2\times10^{6}\,\msun$. 
}
\label{fig:magrest}
\end{figure}

Figure \ref{fig:magrest} displays the rest-frame magnitudes at 1500 \AA\ and for the $r,\, g$~and $V$~filter passbands for the M05 SSP models,  shown for $t<100$~Myr and for a total stellar mass of $2\times10^{6}$. 
In the UV the models are brightest at an age of 3 Myr, when the most massive stars start to die, whereas  in optical bands they are brightest at slightly later times due to the appearance of red supergiants.
After reaching the brightest luminosity $\sim 3-10$ Myr since formation (depending on wavelength), all models  fade monotonically in all considered bands. Hence, catching them within the first $\sim 10$ Myr  since formation gives the best chance to observe the progenitors of present-day globular clusters. 
For instance, comparable UV luminosities (between $M_{1500}=-17$ and $-15$) have been derived for candidate GCPs with similar masses ($1-20 \times 10^6 M_\odot$)
found by \cite{vanzella17a, vanzella17b, vanzella19}, and for  tiny lensed sources identified by \cite{bouwens17} at $z\sim6-8$.
 \subsection{Redshit evolution of population models}\label{sed:magz}
The observer-frame properties of the models in the NIRCam filters are calculated by red-shifting the rest-frame model SEDs to a family of redshifts (from $z=10$~down to $z=3$, in steps of $\Delta~z=0.1$) and for a series of times since formation, hence redshift at which they are observed. The cosmological dimming is calculated using the Flake code ({\it Flexible-k-and-evolutionary- correction}, C. Maraston, {\it in prep.}). For an assumed cosmology, the procedure, calculates the observed-frame magnitudes in arbitrary photometric filters for all model ages and redshifts, including $z=0$. A (large) table of possible evolutionary paths is output, which besides providing the observed-frame and absolute magnitudes of objects with arbitrary ages and star formation histories, it also allows a quick evaluation of the $K$-correction in various filters without the need to approximate. Moreover, as all model ages are considered at all redshifts, there is no need to assume one specific formation redshift in order to follow the evolution, as any choice for this parameter is possible. In this work we shall experiment with a set of formation redshifts and other parameters, as described in the following sections.   
Table 1 provides an example of the observer-frame magnitudes in all JWST filters at different redshift and ages for our typical GCPs of mass $2\times 10^6$ $M_\odot$~based on the adopted SSP models. In particular we list observer-frame magnitudes near the brighest phase, i.e., at age of $10$ Myr since its formation, and hence formed at a slightly higher redshift. The full table can be found here\footnote{
https://sites.google.com/inaf.it/pozzetti-gcps/home}
Note that these magnitudes do not yet include the high-$z$  absorption by the intervening  intergalactic medium  (IGM).

  \begin{figure*}
\includegraphics[angle=-90,width=160mm ]{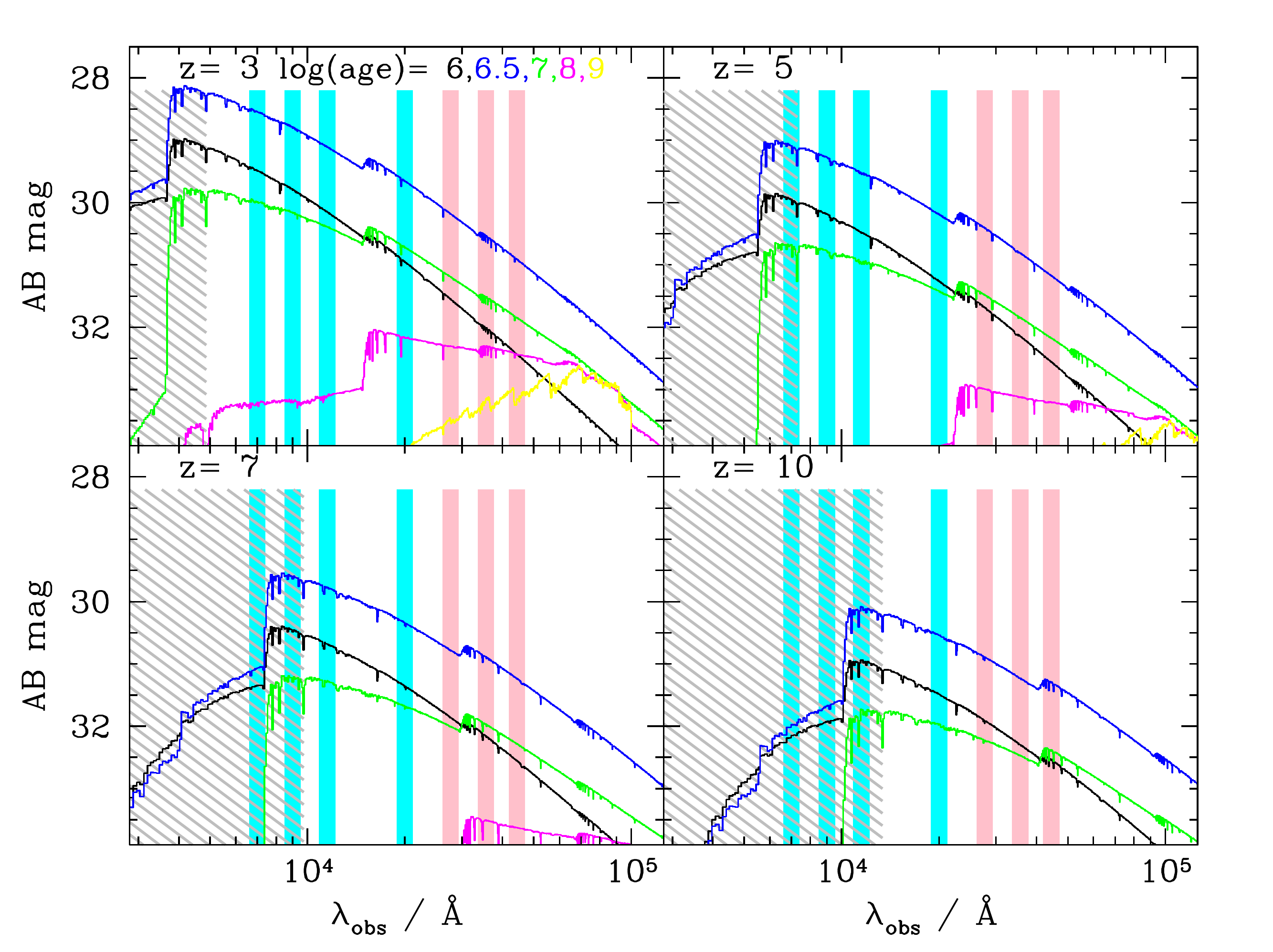}
\caption{Expected observer-frame spectra for our typical GCP of mass $2 \times 10^6$ $M_\odot$,  at different ages (from $10^6$ yr to $10^9$ yr) and at different redshifts (at $z=3$, 5 ,7 , 10 in the 4 panels). Fluxes are expressed in AB magnitudes. Vertical bands reproduce the JWST filters from the two short and long-wavelength channels (in blue and pinkish,  respectively). In the grey shaded region  spectra are affected by  high-$z$ IGM absorption.}
\label{fig:sed_z}
\end{figure*}
  
\section{Colours and luminosities of young SSP{\MakeLowercase{s}} as seen by JWST}
\label{sec:magz}
In this section we make predictions on the detectability of GCPs by JWST, 
using NIRCam imaging under the assumptions mentioned above.  
NIRCam offers high sensitivity imaging\footnote{https://jwst-docs.stsci.edu/display/JTI/NIRCam+Sensitivity}
from 0.6 to 5.0 $\mu$m in 8 broad-band filters (F070W, F090W, F115W, F150W, F200W, F277W, F356W F444W) and consists of two modules pointing to adjacent fields of view, separated by 4.4 arcsec.  Each module observes simultaneously in a short wavelength channel (0.6 to 2.3 $\mu$m) and in  a long wavelength channel (2.4 to 5.0 $\mu$m). The total field of view (FoV) of each NIRCam pointing is 9.7 arcmin$^2$.

\begin{figure}
\includegraphics[width=84mm]{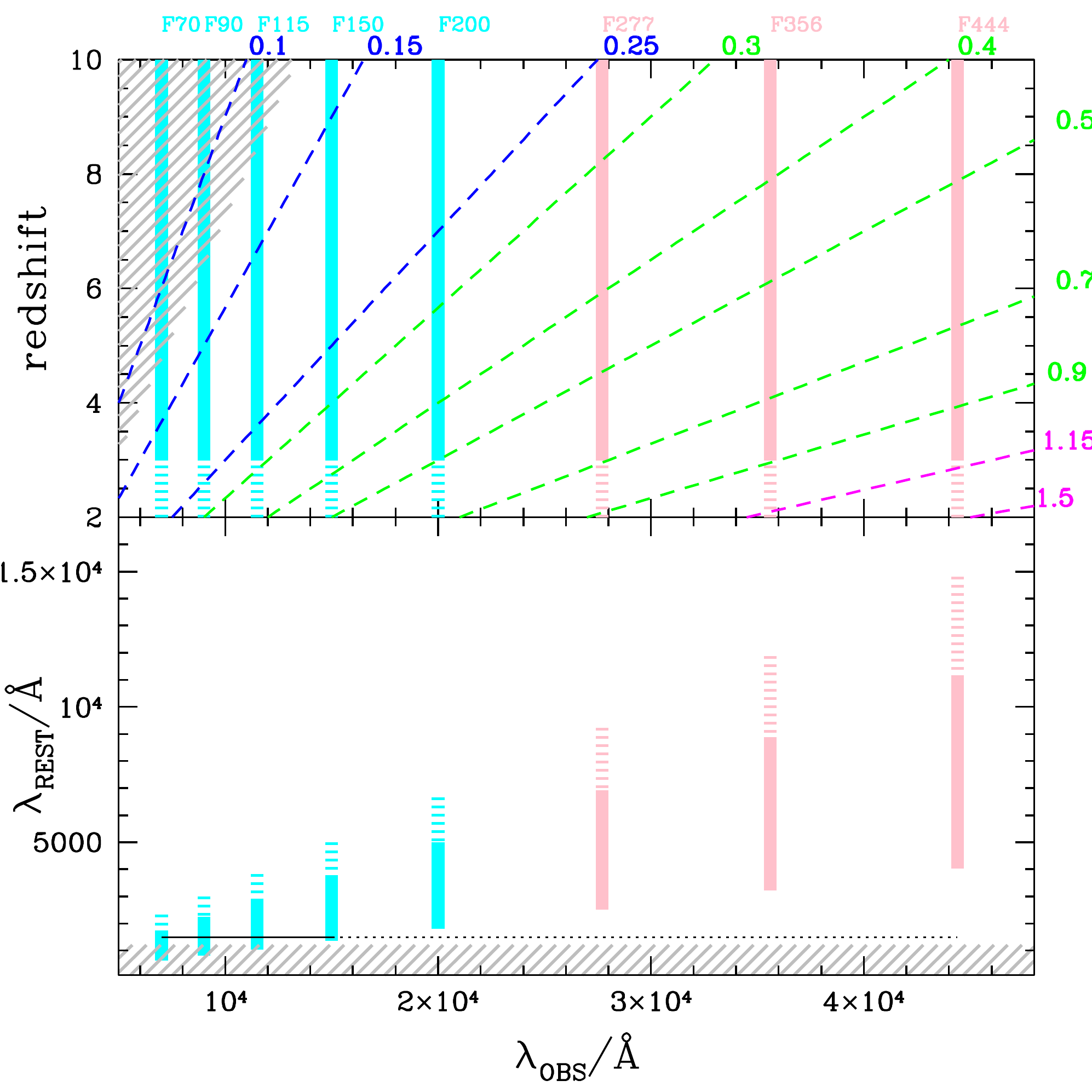}
\caption{The rest-frame wavelength sampled by NirCam filters  for objects at different redshifts. {\it Top panel:}  The dashed lines labeled by the rest-frame wavelength (in $\mu$m) show the corresponding observed wavelength as a function of redshift. The vertical bands give the central wavelength of the NIRCam filters, as in Figure \ref{fig:sed_z}. Thus, the intercepts of the dashed lines  with the vertical bars give the redshifts  at which a given rest-frame wavelength will be sampled by the various NIRCam passbands. {\it Bottom panel:} the vertical bars give the full rest-frame wavelength range sampled by the NIRCam filters for objects in the redshift range $2<z<10$ (dashed bands for $2<z<3$, continuous bands in the redshift  range
 ($3 <z <10$).}
\label{fig:lambda_z}
\end{figure}

Using the stellar population models presented in the previous section, we show in Figure \ref{fig:sed_z} the observed spectra 
for a GCP/SSP of mass $2 \times 10^6$ $M_\odot$,
at different ages and redshifts from $z=3$ to $z=10$, from young ($10^6$ yr) to old ($10^9$ yr) ages. We focus to those on the youngest ages ($\sim10^{6.5}$ yr) as they correspond to the brightest phase of GCPs, hence with the highest chance of being detected. Indeed, after the brightest phase, the flux drops quite rapidly by at least 1 magnitude in a time scale of few Myr at all wavelenghts and redshifts. Notice that for such young ages the spectrum longward of the Lyman break is well represented by a power law with $F_\lambda\sim \lambda^{\beta}$
with only a mild evolution  during the first $\sim 10$ Myr from $\beta\simeq -3$ to $-2.5$. 
Steep UV spectral slopes are indeed ubiquitous among very high redshift galaxies, getting steeper with decreasing luminosity and approaching $\beta \sim -3$ in the luminosity range expected for GCPs (see Figure 1 in \citealt{bouwens14}). Such steep spectral slopes are
also very similar to those  observed in candidate GCPs (\citealt{vanzella16,vanzella19}). This steep UV slope of young SSPs plays a critical role in determining the predicted luminosities and colours of detectable GCPs as a function of their formation redshift.
For this reason, in the reddest channels the flux in the brightest phase is not dramatically lower at $z=10$ compared to $z=3$. 

 Figure \ref{fig:sed_z} illustrates how  NIRCam filters sample different spectral ranges depending on redshift, from UV to optical going from bluest to reddest filters and from lower to higher redshifts. At $z>5$ the bluest filters cover a range of wavelengths shorter than Lyman break and Ly-$\alpha$, which are affected by the absorption by the high-$z$ hydrogen in the IGM  intervening along the line of sight. At $z=7$  this effect is important for the F070W, and F090W passbands and at $z=10$ also for the F115W passband, hence GCPs at such redshift will appear as drop-out objects in those bands. As a consequence, at $z\ge 7$ the GCPs are detectable only in the complementary longer wavelength channels.

\begin{figure*}
\includegraphics[width=84mm ]{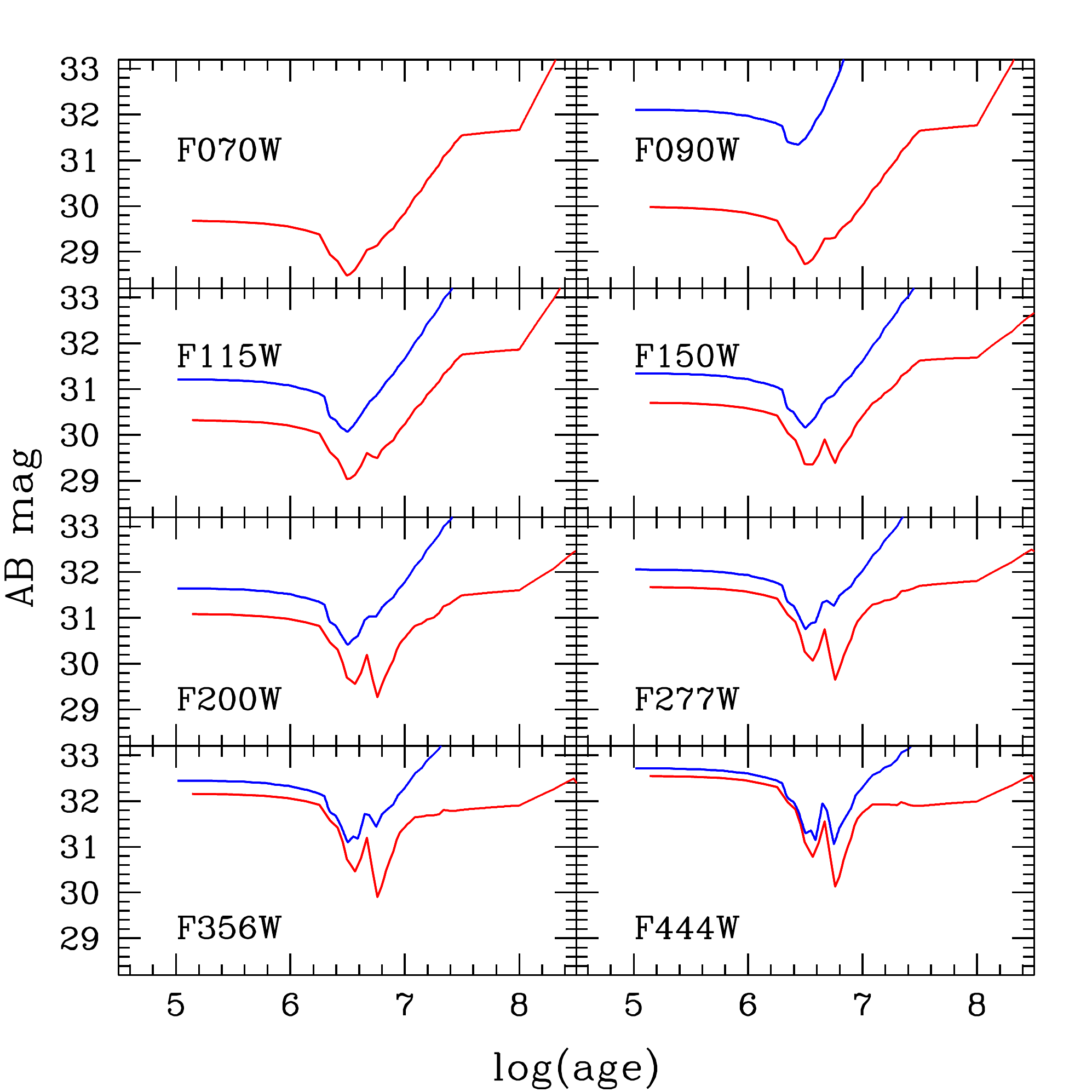}
\includegraphics[width=84mm]{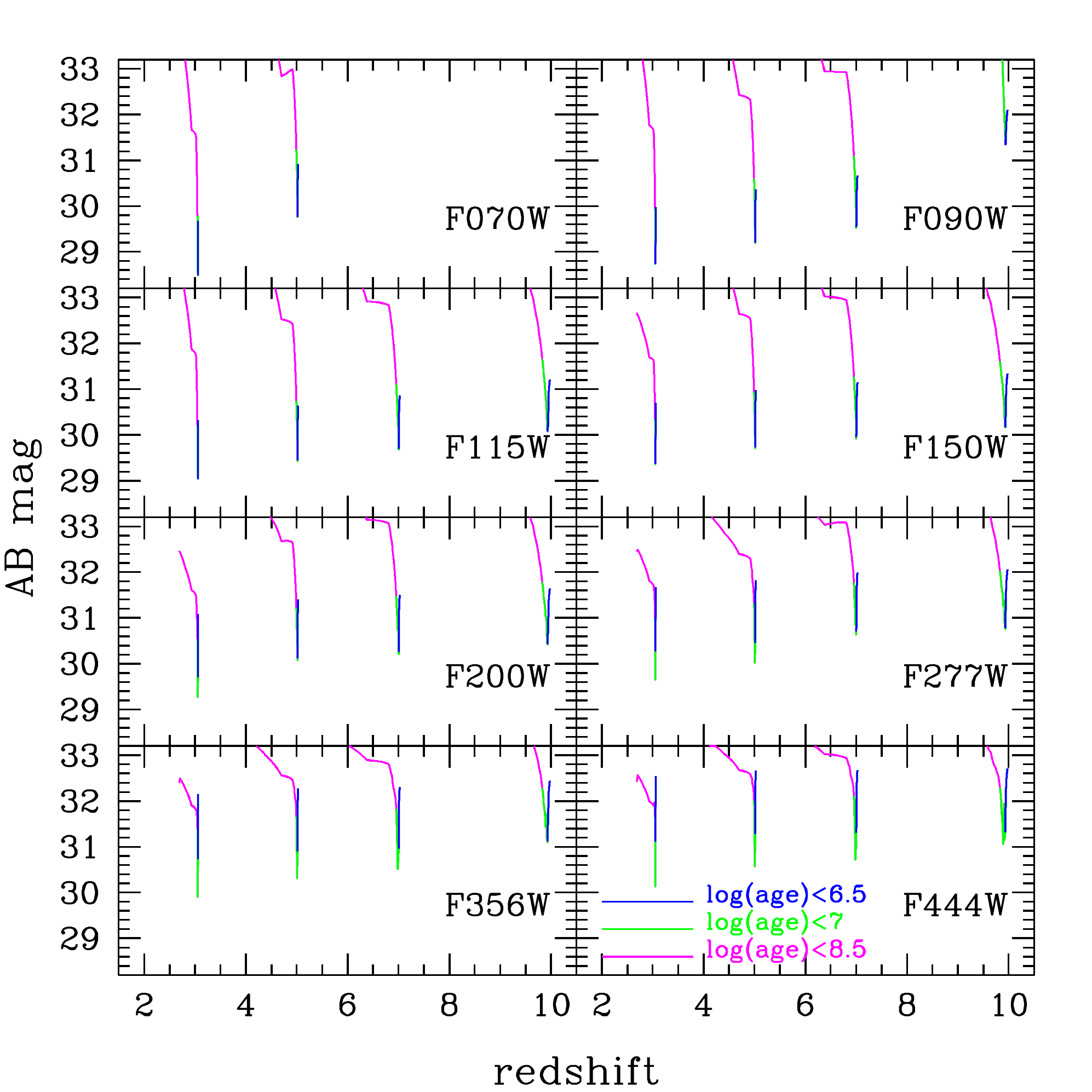}
\caption{Observed NIRCam magnitudes expected for a typical GCP of mass $2 \times 10^6$ $M_\odot$, adopting different formation redshifts.
Different panels show different NIRCam bands.
{\it Left panels} show magnitudes as a function of time since formation, for a GCP formed at  $\zf=3$ (in red) and $\zf=10$ (in blue).
 {\it Right panels} show magnitudes as a function of redshift for $\zf=3, 5, 7, 10$. Different colours refer to different ages after the formation, as encoded in the bottom-right panel. } 
\label{fig:mag}
\end{figure*}

To guide the eye, we show in Figure \ref{fig:lambda_z} the rest-frame wavelengths as sampled  by the various NIRCam  filters at redshifts in the range $2<z<10$,  and the corresponding rest-frame spectral ranges. In particular, the short-wavelength channel samples the UV rest-frame at $z>5$. For example, in the range $3<z<10$ the filter F200W observes the rest-frame from $\lambda_{\rm REST} =1800$ to 5000 \AA.  Conversely, in the same range  the rest-frame at $1500$ \AA\ will be observed only by filters bluer than F150W. The long-wavelength camera instead would cover almost exclusively the rest-frame optical range ($2500<\lambda_{\rm REST}<10,000$  \AA), even at $z=10$.
Furthermore, as already pointed out, the effect of the IGM absorption, dropping-out the object from bluer passbands, is important only for F070W, F090W and F115W and for $z>5$, 6.5 and 8.5, respectively. 

Using the table of redshifted magnitudes described in Section \ref{sed:magz}, we derive the expected  fluxes/magnitudes in the various NIRCam passbands as a function of age and redshift/epoch  of formation ($\zf$ and $\tf$, hereafter), to which we add 
 the IGM absorption effect by attenuating the resulting flux/magnitudes using  the prescriptions of \cite{madau95}.
Figure \ref{fig:mag}  show the expected magnitudes, as a function of time since formation and as a function of redshift, for representative formation redshifts in our interval ($\zf=3,5,7,10$).
Adopting a typical GCP with mass log$(M^*/M_\odot)$=6.3 
we find that the maximum fluxes range between 29 and 31 in magnitude, depending on the filter. 
As already mentioned, the brightest phase is very fast (few Myr) and peaks at very young ages ($\sim 10^{6.5}$ yr), hence it lies at redshifts very close to $\zf$. 
For ages older than 10-100 Myr,  GCPs fade by several magnitudes (from 2 to 3, i.e., a factor 5 to 10 in flux). 
Note that the bright phase is  always short, indipendently on the formation redshift and on the filter, lasting few Myrs  before dropping by 1 magnitude, or at most up to 100 Myr to fade by 2 magnitudes for low formation redshift ($\zf=3$) and reddest filters, 
but always covering a broder redshift range for higher formation redshift.
We also note that the brightest fluxes are expected at the lowest redshift and in the bluest filters, given that the maximum flux is in the UV rest-frame (see previous section). At  higher redshift and redder filters we expect GCPs to be fainter  due to, respectively, higher distance and filters sampling an intrinsically fainter part of the  SED. 
However, for filters redder than F150W the difference in the maximum fluxes between different $\zf$ is always less than $\sim 1$ magnitude. Actually, as evident from Figure \ref{fig:mag}, during the first 10 Myr since formation the observed magnitude range spanned by GCPs only sligthly change with the 
formation redshifts. This is a consequence of the near power-law shape of the spectrum of young SSPs (see again Figure \ref{fig:sed_z}), hence a negative $k$ correction largely compensates for the increasing distance with redshift.

\begin{figure}
\includegraphics[width=84mm]{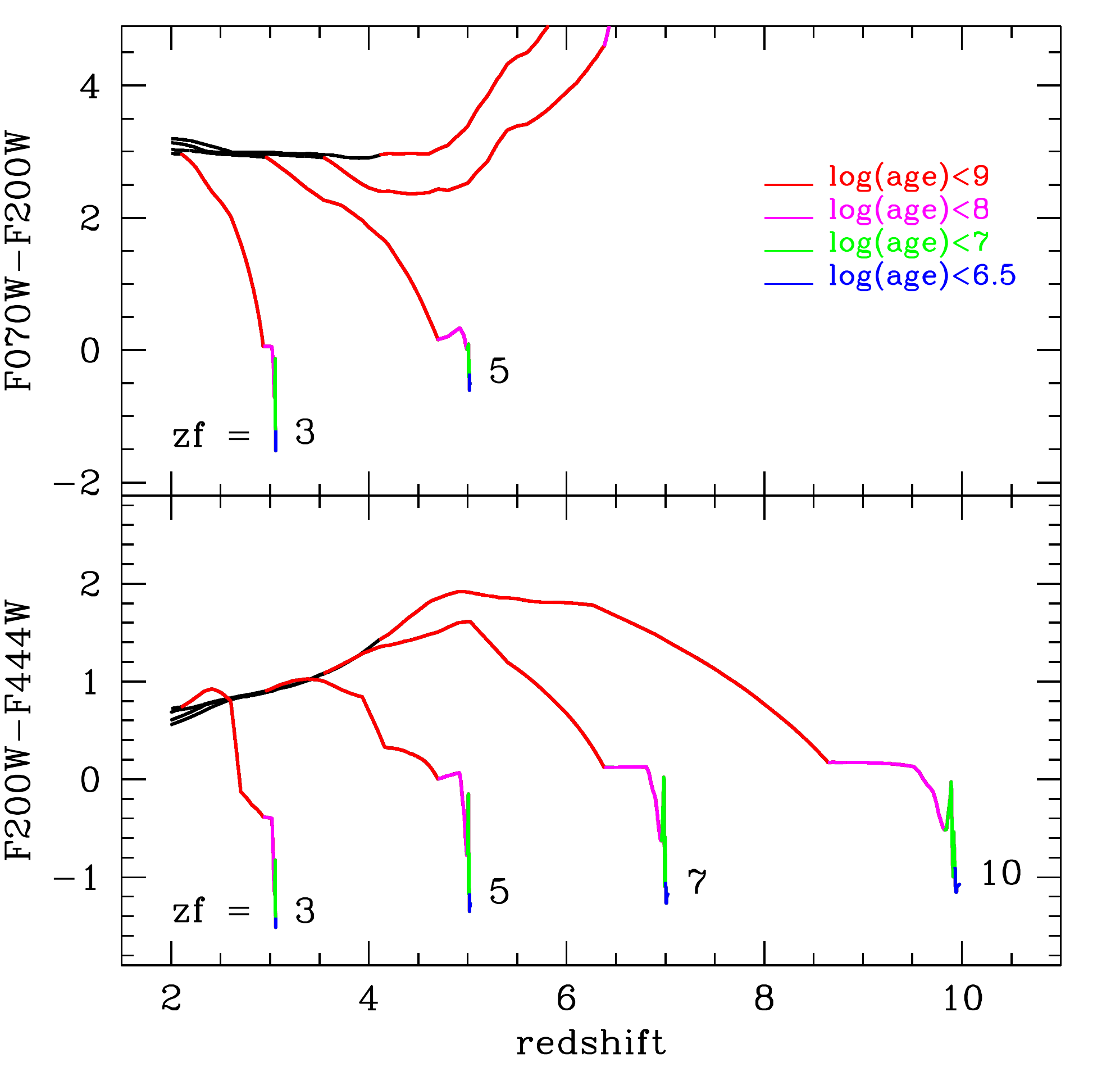}
\caption{Observed F070W-F200W and F200W-F444W colour evolution of GCPs forming at redshifts $\zf=3, 5, 7, 10$. Different colours refer to age ranges as indicated
in the insert. For $\zf=7,$ and 10 GCPs would appear as F070W dropouts.}
\label{fig:colors}
\end{figure}

As a further example, Figure \ref{fig:colors}   shows the expected  F070W-F200W and F200W-F444W colours as a function of redshift for the same $\zf$  values.  In both cases the bluest colours are found at $z\sim\zf$, i.e., at formation and shortly thereafter. The  F070W-F200W colours become very red at $z>5$, due the effect of IGM absorption in the bluest of the two filters. Also for the colours, as for magnitudes, there are no big differences for different $\zf$. In particular, during the first 10 Myr  the F200W-F444W colours
are quite insensitive to $\zf$, which is again due to the power-law shape of the SED. In conclusion, the  shape of the GCP spectrum during the first $\sim 10$ Myr since formation has the  interesting effect that both luminosities and colours are quite insensitive to the formation redshift, unless photons that would be detected in a given passband suffer from IGM absorption. Actually, the only way of measuring a photometric redshift for GCP candidates will be through  the dropout technique, as is currently the case for very high redshift galaxies.

\begin{figure}
\includegraphics[width=84mm ]{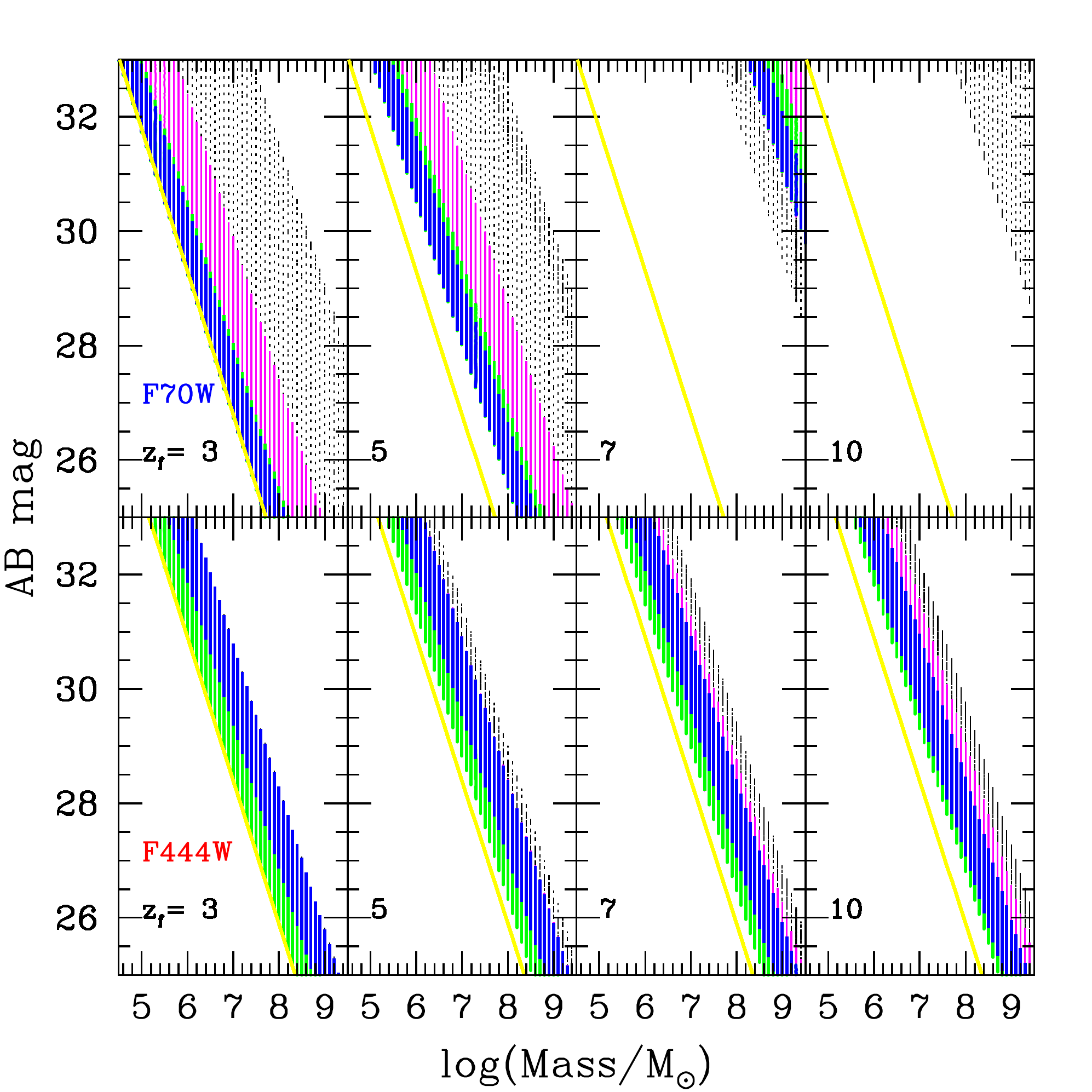}
\caption{Observed magnitude-mass relation for a GCP, adopting different formation redshifts ($\zf=3, 5, 7, 10$). Different colours refer to different ages after the formation, as in Fig. \ref{fig:mag}, and in grey for ages greater than 1 Gyr. Top panels for the F70W filter, bottom panels for F444W.
The yellow line is at a fixed position in all panels of a given filter.} 
\label{fig:mag_mass}
\end{figure}

Finally,  we illustrate the relation between magnitude and mass in the different filters and for various formation redshifts. As  luminosity scales linearly with mass, their relation can be written as:
\begin {equation}
{\rm log}(M/M_\odot)=0.4*(F_0-F)+{\rm log}(M_0/M_\odot)
\end{equation}
where $F_0$ is the magnitude in a generic filter at a given reference mass $M_0$.
In Figure \ref{fig:mag_mass} we show that this relation has a minimum in the bright phase at young age ($<10$ Myr) and how it depends on age and $\zf$. For the bluest filter (F70W) there is a strong dependence on $\zf$ due to the effect of high-$z$ IGM  on the UV flux.  Instead, for the reddest filter (F444W) this relation depends only mildly on $\zf$  (less than 1 magnitude, or less than a factor 2 in mass) and on age. In each set of panels the yellow line coincides with the left edge of the band in the lowest redshift panel and for reference it is then reproduced identically in the other three panels. This helps to appreciate how little these magnitude-mass relations depend on formation redshift. Similar, self-explanatory  figures for  other passbands can be found in the Appendix. 
From these relations we can define the minimum mass for a GCP to be observed at a given magnitude in a given filter.
For the representative mass ${\rm log}(M_0/M_\odot)=6.3$ the corresponding magnitudes in the equation above are, for $\zf=3$, $F_0$ = 28.55, 28.75, 29.05, 29.35, 29.25, 29.65, 29.85 and 30.15 for the various NIRCam  filters, respectively from the blue to the red. Similarly, for a given observed magnitude $F=30$, the minimum corresponding masses, for $\zf=3$, can be derived from the previous equation and are about log$M/\msun$ = 5.7, 5.8, 5.9, 6.0, 6.0, 6.1, 6.2 and 6.3 for the various NIRCam filters, from the bluest to the reddest one, respectively. These minimum masses increase with increasing $\zf$, but only slightly for long-wavelenght passbands (see Figure \ref{fig:mag_mass}).
Thus, reaching magnitude 30 in all bands the minimum detectable GCP mass increases with wavelength, but only by a factor $\sim 4$ from the bluest to the reddest NIRCam filter.

For comparison candidate GCPs at $3\lsim z\lsim 6$ in \cite{vanzella17a} and \cite{vanzella19} are as bright as $-17 \lsim  M_{1500} \lsim -15$, or $29\lsim m_{\rm F105W}\lsim 32$,  and derived masses in the range $\sim 5\times 10^5$  and $\sim 10^7\,\msun$ and ages less than 10 Myr.

\section{Predicted number counts of  Globular Clusters Precursors}

Forecasts for any future survey require as input the luminosity/mass function in order to determine the number of objects above a given sensitivity.   
Indeeed, in a homogeneous and isotropic universe, the number of objects of each type (GCPs in our case) brighter than a given magnitude $m_{\lambda}$ can be calculated from the integral:
\begin{equation}
	{N_{\zf} (< m_\lambda)} = \!\!\!\int_{\zmin}^{\zf} \int_{M_{\rm min} (m_\lambda, z, \zf)}^\infty  {dN\over d{{\rm log}}M} \,\frac{dV}{dz} \, d{\rm log}M\,  {dz}\;\;\;
\label{eq:numbers}
\end{equation}
\noindent where $\zf$ is the adopted redshift of formation for our GCPs and ${dN\over d{\rm log}M}$ (function of $M$)  is the GCPs mass distribution from Equation (1),
with log$(M^*)=6.3$, normalized to include all GCPs, i.e. $N_\circ = 1.15$  Mpc$^{-3}$.
We assume that our GCPs have concluded  their primordial phase at z$_{\rm min}$=2. This assumption does not affect dramatically our computation, since their luminous and mass loss phases are even shorter than 1.5 Gyr (the maximum time elapses between z=2 and the maximum redshift of formation assumed, $\zf=10$).
The integration over $d{\rm log}M$ extends from ${M}_{\text{min}}$ which depends on redshift, age, $\zf$ and magnitude limit ($m_\lambda$) in a given band and it can be derived from our model SSPs  and take into account the aging of the GCP population.
In practice, we invert  the relation between magnitude and mass using Table 1. The minimum mass ${M}_{\text{min}}$ in Equation (1) is then derived from the equation:
\begin{equation}
{\rm log}M_{\text{min}}=-0.4 \; ( m_\lambda -  m_\lambda^{6.3}(z,\zf) )+6.3
\end{equation}
\noindent 
where $m_\lambda^{6.3}$ is the magnitude observed in a given filter at $\lambda$ for a GCP of mass log$(M/\msun)=6.3$, at a given redshift ($z$)  taking into account the evolution in time since formation ($t$) for any given formation redshift ($\zf$). This has been derived from the intrinsic evolving spectra as described in Section \ref{sec:magz} and
further attenuated by the IGM.

\begin{figure*}
\includegraphics[width=150mm ]{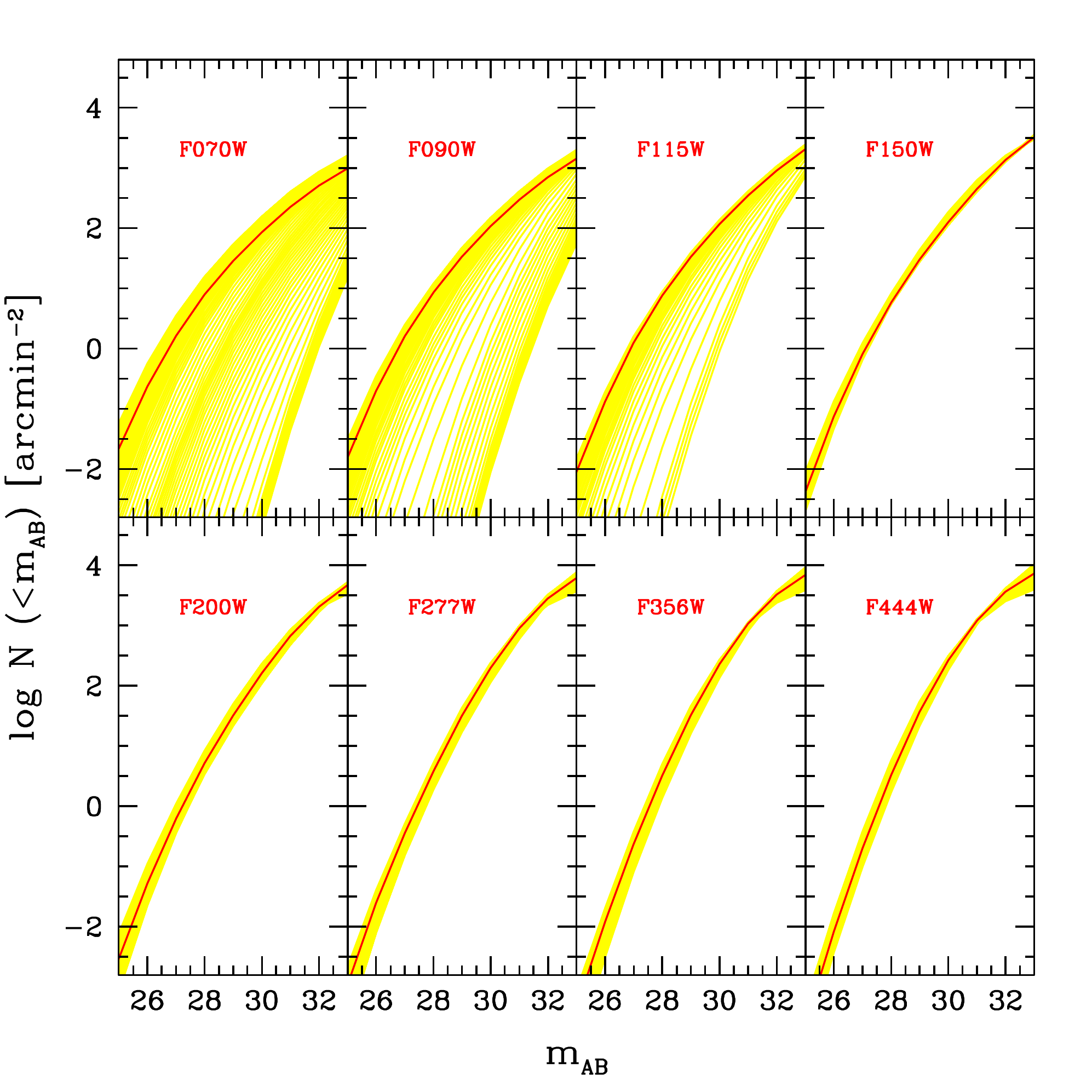}
\caption{Expected GCPs cumulative counts in all NIRCam bands as from Eq. (5) for log$(M^*)=6.3$. Each panel refers to a different NIRCam  band. In yellow the number counts for various formation redshifts in the range $\zf=3 - 10$, assuming that all GCPs form at a given $\zf$. The red lines show the total counts adopting a continuous formation redshifts (constant in time) within the same redshift interval. }
\label{fig:counts_all}
\end{figure*}

Finally, assuming  that the bulk of GCs are formed  
in the range $3<\zf<10$, 
corresponding to a lookback time in the range of $t_{\rm lb}=11.5-13$ Gyr, 
then the total number densities of GCPs `brighter than $m_\lambda$ is given by:
\begin{equation}
\begin{tabular}{ r l }
\({N_{\rm tot} (< m_\lambda) }  $ & \(=\)
\\
\\
& \hspace{-2.2 truecm} = \(\mbox{\LARGE\(\!\!\int_{t_{\rm lb}^{\rm inf}}^{t_{\rm lb}^{\rm sup}}\!\!\!\int_{z_{\rm min}}^{z_{\rm f}(t_{\rm lb})} \!\! \int_{M_{\text{min}} {\small (m_\lambda, {z}, {z_{\rm f}})}}^\infty  \)}{\cal F}(t_{\rm lb})\LARGE{{dN\over d{\rm log}M} \, \frac{dV}{dz}}\, d{\rm log}M\, {dz}  \, {dt_{\rm lb}},\)
$$
\end{tabular}
\label{eq:counts}
\end{equation}

\noindent where ${\cal F}(t_{lb})$ is the fraction of globular clusters produced per unit time. Here we assume it constant in time (not in redshift) and therefore ${\cal F}(t_{lb})= 1/(t_{\rm lb}^{\rm inf}- t_{\rm lb}^{\rm sup})$.

From Equations \ref{eq:numbers} and \ref{eq:counts} we obtain the redshift distribution per unit redshift (${dN\over dz} (< \text{m}_\lambda; z, z+dz)$ by integrating over the specific redshift range ($z, z+dz$).
We stress here that at all observed magnitudes, all masses, formation redshifts and times since formation can contribute to the number counts.

The cumulative number densities per arcmin$^2$ predicted by our model are shown in Fig. \ref{fig:counts_all} in the various NIRCAm filters, from the bluest (F070W) to the reddest (F444W) and for our adopted Mass Function with log$(M^*/M_\odot)=6.3$, 
assuming a birth rate constant in time in the redshift range $\zf=3 - 10$. We also show the counts adopting a single $\zf$, i.e. all GCPs form at the same $\zf$.
The corresponding tables with cumulative and differential counts can be found at: $https://sites.google.com/inaf.it/pozzetti-gcps/home$.
We note that the predicted counts are fairly insensitive to the formation redshift, 
with the exception of the three bluest passbands, because of the dropout effect. 
The similarity of the predicted counts per unit area (not per unit volume) for different formation redshifts is due to a combination of effects. First, as already discussed, the magnitudes/fluxes during the brightest phase are fairly insensitive to the formation redshift, being at most 1 magnitude fainter for $\zf=10$ compared to $\zf=3$.
Furthermore, even if the duration of the bright phase is similar for different $\zf$ in term of time, it is always broader in term of redshift range for high $\zf$. This will end in a larger volume per unit area for high $\zf$ GCPs, which compensates for the slightly fainter fluxes, hence determining similar effective counts per unit area.

\begin{figure}
\includegraphics[width=84mm ]{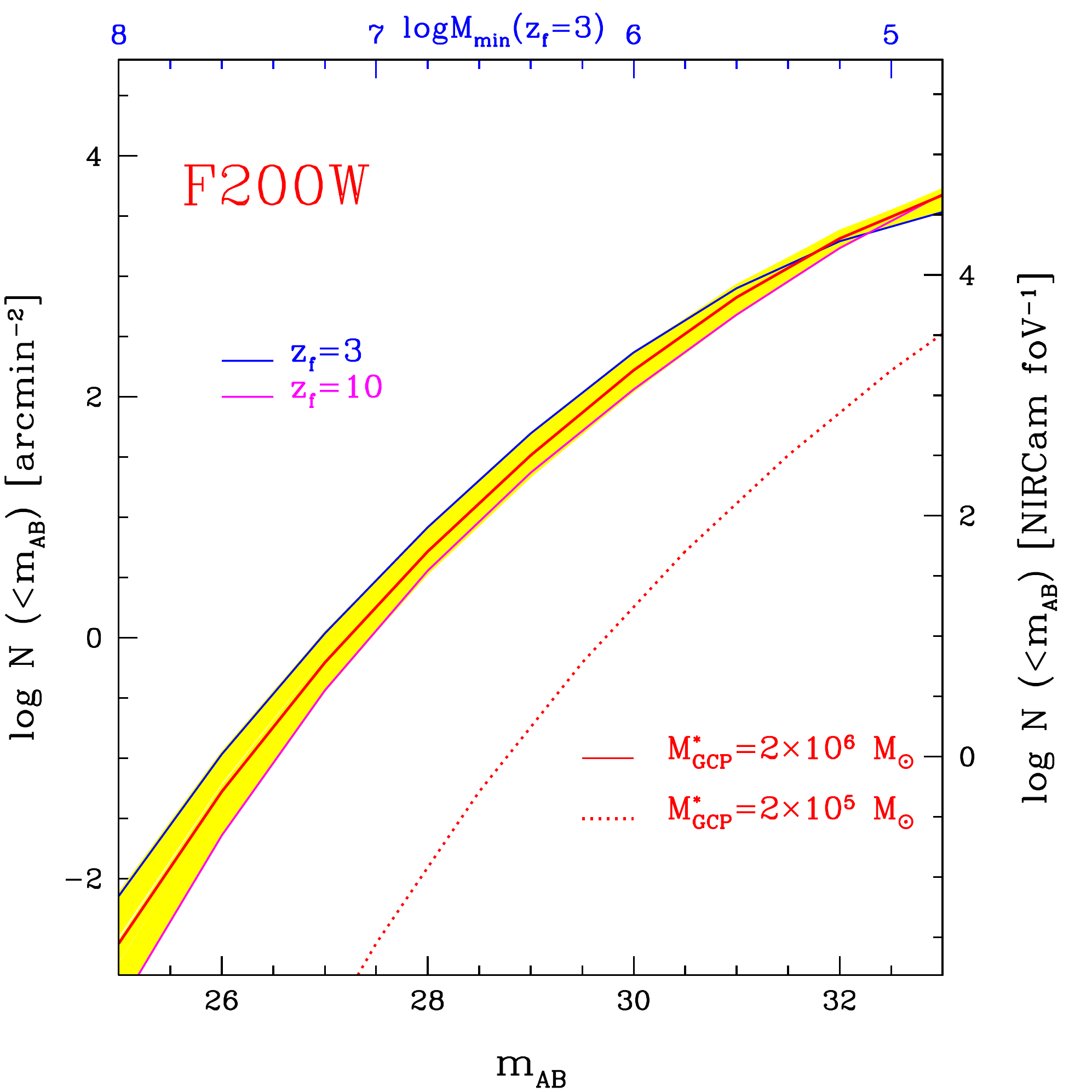}
\caption{Expected GCPs cumulative counts per  arcmin$^2$ (also reported per NIRCAm FoV on right y-axis) for the F200W band, replicated from Fig.~\ref{fig:counts_all}. The dotted red line shows the cumulative counts assuming that GCPs formed with the same mass function of present day GC (i.e., log$ M^*/M_\odot=5.3$).
In yellow the number densities for various formation redshift, of which  in blue for $\zf=3$ and in magenta for $\zf=10$.
}
\label{fig:counts_F200}
\end{figure}

\begin{figure}
\includegraphics[width=84mm ]{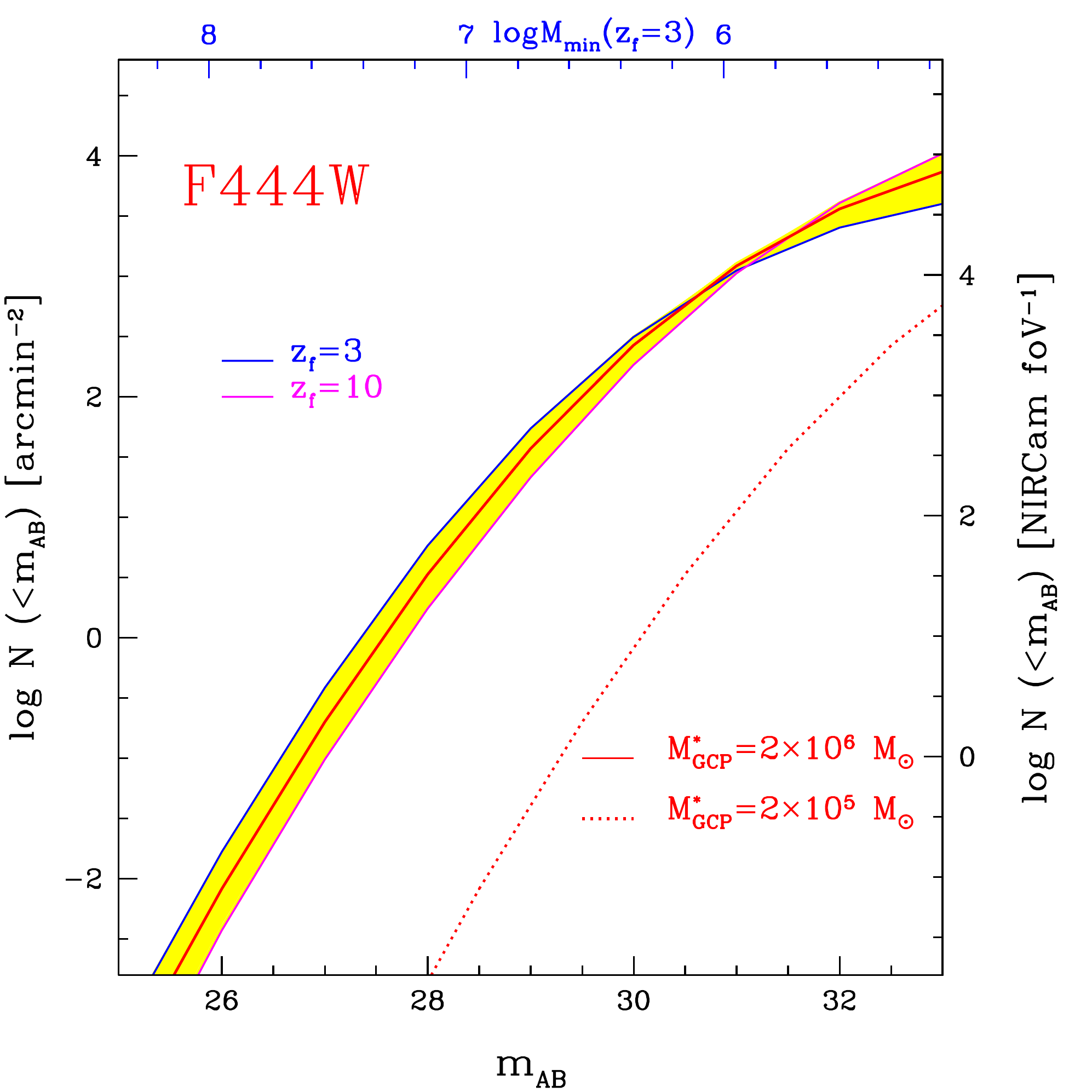}
\caption{The same as  in Fig. \ref{fig:counts_F200}, for the F444W band.
}
\label{fig:counts_F444}
\end{figure}
  
\section{Perspectives at detecting GC{\MakeLowercase{s}}  in formation, with caveats}

We show in Figures \ref{fig:counts_F200} and \ref{fig:counts_F444}, in particular,  the same for the F200W and F444W passbands. Notice that the upper scale gives the minimum mass for a GCP (which form at $\zf=3$) being as bright at its peak luminosity as indicated by the lower scale. For example, only GCPs more massive than $\sim 10^7\,\msun$ could be brighter than mag=28 and only those more massive  than $10^6\,\msun$ could shine brighter than mag=30. 
At these masses, as shown in Figure \ref{fig:magrest}, we expect, for instance for a GCP of $\sim 2 \times 10^6\, \msun$ to be as bright as $M_{1500} \sim -17$ at its peak.
The upper and lower borders in the counts, with single $\zf$, correspond to all GCPs forming at $z=3$ or at $z=10$, respectively, with the exception of the very faint magnitudes case, in particular for F444W passband, where there is an inversion with those formed at the highest redshift starting to dominate the counts. This effect is mainly due to the fact that GCPs formed at $\zf=3$ are brighter (by 1 magnitude) and therefore they reach the maximum density at the peak of the Gaussian Mass Function (at $logM/M_\odot =6.3$ and magnitude $\sim 30.5$) and thereafter, at magnitudes brighter than those of GCPs formed at zf=10, start to diminish in density relative to them.
The figures also include  the expected number counts for log$(M^*/M_\odot)=5.3$, i.e., assuming GCPs formed with the same mass function of present day GCs, as if they had  suffered no mass loss at all. This is clearly a strict lower limit to the expected number counts. To the extent to which an $M^*$ ten times higher than that can be regarded as an upper limit, then we expect that the real counts will fall somewhere in between the dotted line and the yellow band. Note that the vertical scale in these two figures gives the number counts per NIRCam FoV, hence for log$M^*/\msun =6.3$ one expects NIRCam to detect of order of $\sim 1,000$ GCPs down to mag =30 in either the F200W or the F444W passbands. In the most conservative case, this number falls down  to $\sim 10$ detections per NIRCam pointing. So, in conclusion, one expects from $\sim 10$ to $\sim 1,000$ GCP detections, depending on the actual value of the "mass budget factor" ({\it mbf}) in the range 1 to 10, that future NIRCam observations will actually allow us to estimate. 
The first opportunity to check these numbers will be offered by the JWST Early Release Science (ERS) observations that will include the coverage of $\sim 100$ arcmin$^2$ with NIRCam\footnote{{\small https://jwst.stsci.edu/observing-programs/program-information?id=1345}} in the five reddest band down to mag $\sim 29$ in the F200W band (28.6 in the F444W band). 
We estimate that $\sim 3,700 ~(1,400)$ candidate GCPs 
should be detected in the F200W (F444W) band during ERS for {\it mbf}=10, which are
drastically reduced to less than $20$  
for {\it mbf}=1.
Then ERS observations will be followed by  the NIRCam guaranteed time observations (GTO)\footnote{https://jwst-docs.stsci.edu/display/JSP/JWST+GTO+Observation+Specifications} planned to reach mag=29.8 (at $10\sigma$ for point sources) over an area of 46 arcmin$^2$. At this limiting magnitude, objects more massive than $\sim 10^6\,\msun$ should be detected while near maximum light, providing up to $\sim 5,500 ~(8,300)$ objects in the F200W (F444W) band for {\it mbf}=10, reduced to just $\sim 55$ ($20$) detections for {\it mbf}=1.
Moreover, the GTO team plans also to reach mag=28.8 over an area of 190 arcmin$^2$, which from Figure \ref{fig:counts_all} corresponds to detecting $\sim 20$ or $\sim 4,000$ GCPs, for a mass budget factor 1 or 10, respectively.
Combining together the ERS data and the GTO deep and broad observations, NIRCam should detect from $\sim 100$ to $\sim 14,000$ GCPs, respectively for {\it mbf}=1 and 10. But see caveats in the following section. 

\begin{figure}
\includegraphics[width=84mm ]{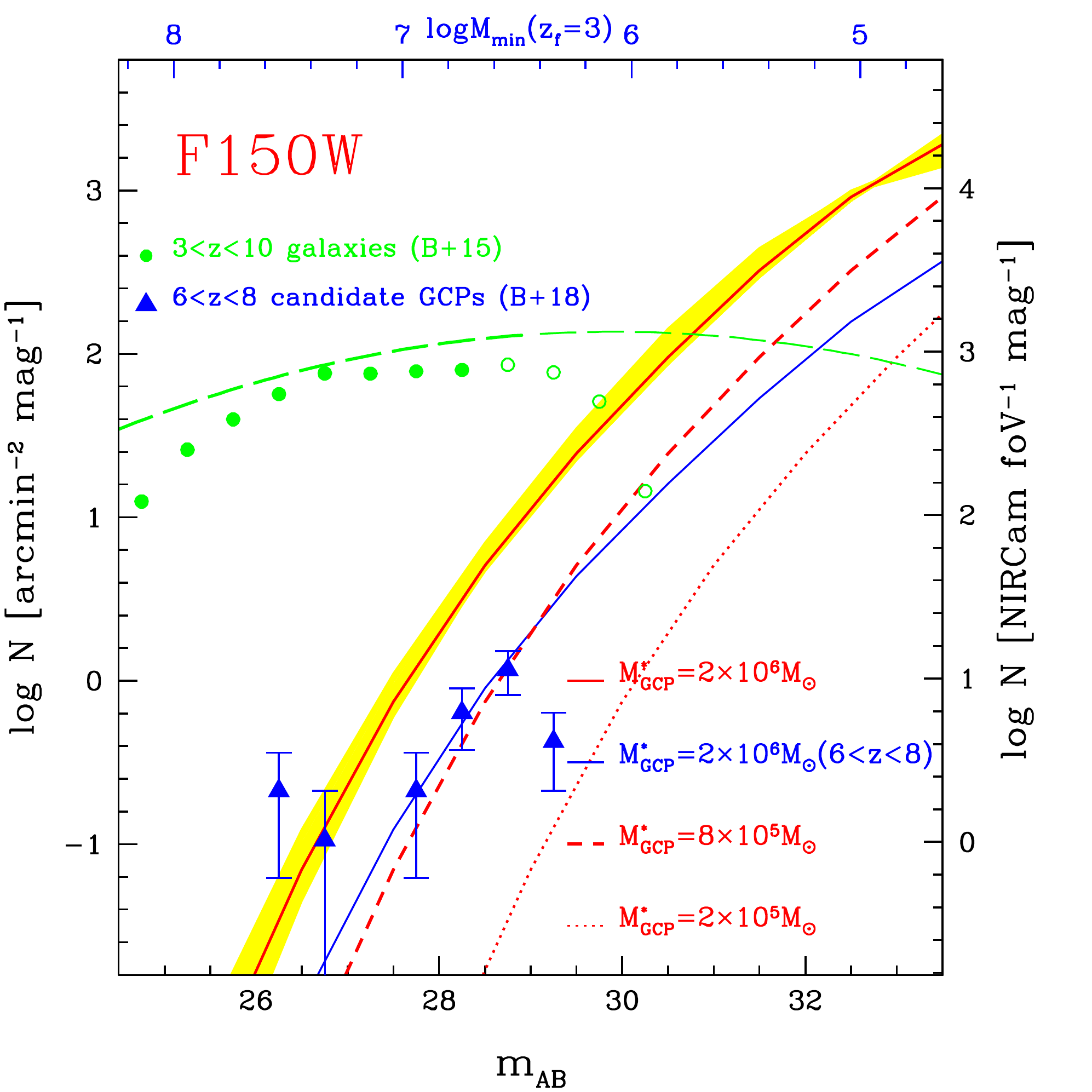}
\caption{The differential number counts for the F150W band. In red the total GCPs counts adopting a continuos formation redshifts ($\zf=3, 10$).
The green dots show the counts for the $3<z<10$ galaxies from Bouwens et al. (2015) in the HST F160W passband. The green long-dashed line shows the total number counts for galaxies at all redshifts \protect\citep{madau00}. In addition, we show the counts of candidate GCPs revealed in the HFF by \protect\cite{bouwens18} in the redshift range $6<z<8$ (blue triangles with error bars). For comparison we show also our model (with $ M^*=2\times 10^6\,\msun$), limited to the same redshift range (blue solid line), and the models adopting $M^*=8\times 10^5\,\msun$ and $2\times 10^5\,\msun$ (dashed and dotted red line, respectively). 
}
\label{fig:counts_F150}
\end{figure}

Moreover, Figure \ref{fig:counts_F150} shows the {\it differential } GCP number counts for the F150W passband. We compare our GCP predictions to 
the current number counts in the HST F160W passband for galaxies in the redshift range 3 to 10 \citep{bouwens15} along with 
the number counts for galaxies at all redshifts, from a compilation of several extragalactic surveys, fitted and extrapolated to unobserved fainter magnitudes \citep{madau00}.
Apparently, for a mass budget factor of 10, GCPs start dominating the high redshift counts just beyond the \cite{bouwens15} limit, and beyond $m_{AB}\simeq 30.5$,  corresponding to a peak luminosity of $M_{1500} \sim -15.5$, could dominate over the total galaxy counts. This transition from galaxy-dominated to GCP-dominated number counts should produce a sharp  inflection in the overall luminosity function from whose location it should be  possible to determine the mass budget factor. 

Finally, the comparison between our number counts predictions and candidates GCPs from \cite{bouwens18}, revealed as compact objects (with size $<$ 40 pc, thanks to lensing effect) by the Hubble Frontier Fields (HFF) program, seems to suggest an encouraging agreement, but better be cautious about its interpretation. Formally, the observed counts of these compact objects at $z=6-8$ suggest that $M^*$ could be at least $8\times 10^5 \msun$, i.e., 4 times more massive than local GCs, or even 10 times ($M^*=2\times 10^6 \msun$) if GCs were to form at a constant rate in the range $z=3-10$, as assumed in our model. In such case, \cite{bouwens18} would have seen only the fraction (20\%) formed in that redshift range $6<z<8$, most likely being all metal poor.
However, not all such objects may eventually end up as GCs $\sim 13$ Gyr later, especially those with masses well above $\sim 10^7\,\msun$, so we cannot claim to have already measured the mass budget factor. Clearly, reaching down to mag=30-31 will be critical in this respect.

All number counts presented in this section can be found on the same site along with the full Table 1 
({\it https://sites.google.com/inaf.it/pozzetti-gcps/home}).

\section{Perspectives at detecting GC{\MakeLowercase{s}}  in formation, with caveats}
\label{sec:caveats}

The number counts presented in the previous section follow from a series of assumptions that may or may not be verified in nature, either in one direction or the other.
One assumption is that GCPs can be described as SSPs. This is obviously at variance with the multiple generation phenomenon, so it appears that real GCs have formed 
in a more or less extended series of individual bursts, interleaved by inactive periods of unknown duration (see Calura et al. in prep., for a model with multiple populations). 
If all star formation activity was confined within less than $\sim 10$ Myr, then our single SSP assumption should not be grossly in error. For longer separations between bursts,  the above number counts should be considered as referring to the brightest event, hence the number counts should be reduced by a factor of $\sim 3$ if such brightest event produced just 1/3 of the final mass of the GCP. In any event, at least in some scenario, the {\it supernova avoidance} 
requirement \citep{renzini15} dictates that all bursts should take place within few Myr, or being separated by more than $\sim 30$ Myr having allowed supernova ejecta do leave the system.

Our modelling  may suffer by another limitation in that it considers GCPs consisting  only of single, non rotating stars. Direct evidence indicates that  the majority  of young massive (O-type) stars are members of binary systems \citep{sana12}. Thus, binary members interactions can affect the resulting SED compared to the SSP approximations, in particular during the first $\sim 10$ Myr since formation, though synthetic stellar populations including massive binaries indicate that the  effects  on  the SED are relatively modest \citep{eldridge17} as is the effect of rotation \citep{leitherer14}.

Another assumption is that GC stars formed following the initial mass function (IMF) proposed by \cite{chabrier03}, which may or may not be the case. Here  we can only say  that massive stars must have formed, given the presence of many pulsars in today's GCs (e.g., \citealt{manchester91}), hence the IMF cannot have been too steep. It cannot  have been too flat either, otherwise  clusters would have dissolved in response to stellar mass loss \citep{chernoff90}. Our exercise  assumes a slope near Salpeter (-2.35) between $\sim 1$ and $\sim 100\;\msun$. If the IMF was (slightly) flatter (steeper) than this, then the number counts presented in the previous section would have been underestimated (overestimated). The effect of varying the IMF is explored in \cite{jerabkova17}.

Then we have assumed that GCPs at high redshift suffered negligible reddening in the rest-frame UV. As already mentioned, this is likely the case for metal poor GCPs, say those with [Fe/H]$\lsim -1.0$, which account for roughly half of the GCs in the local Universe. However, the metal-rich half of the local GC populations must have formed when the build-up of the today hosting galaxy was already quite advanced, given the mass-metallicity relation of high-redshift galaxies \citep{erb06,kashino17}. Within the MW, most metal rich GCs belong to the Galactic bulge and must have formed along with bulge itself, over $\sim 10$ Gyr ago \citep{ortolani95,renzini18}, hence in a chemically enriched and dusty environment, such that GCPs must have suffered substantial UV extinction making unlikely they could be detected.
If so, all predictions made in the previous section should be cut by a factor $\sim 2$.

The physical nature itself of GCPs remains basically unknown. They might have been just compact, somewhat more massive GCs or they may have been the nuclei of dwarf galaxies, as was suggested by \cite{searle78},  most of which later dissolved. Indeed, the Fornax and Sagittarius dwarfs contain an unusually large number of GCs for their mass. In this respect, one question is whether one should consider the whole mass of the dwarf as the mass of the GCP, or just that of the compact object hosted by the dwarf. Following the argument of \cite{elmegreen17}, given the likely $\sim$ kpc size of dwarfs the typical time scale of star formation was of order of $\sim 10^8$ yr rather $10^6$ yr as for GCs, and \cite{zick18} have shown that when forming individual GCs in a Fornax-like precursor would have overshined  the underlying galaxy. Thus, dwarfs hosting forming GCs could not be adequately described by our SSP approximation for GCPs, hence dwarfs parent to GCs  are unlikely to be included in the number counts presented in the previous section, as they would be substantially fainter than massive GCPs younger than $\sim 10$ Myr. In this respect, once a suitable number of candidate GCPs will be found, then stacking them could actually reveal the presence of host dwarfs, and a concrete example has been documented by \cite{vanzella19}. Still, even if not qualifying as GCPs from the observational point of view, dwarfs might have provided nuclearly processed material for the formation of GC multiple stellar generations. Hence, detecting and characterise  the immediate environment of GCPs  should provide critical insight on the process of GC formation. Distinguishing GCPs from their dwarf hosts (or in general from high redshift galaxies) will not be trivial. GC-size objects will appear as point-like in NIRCam images, given their $\sim 200$ pc resolution at these redshifts \citep{renzini17},  and hosting dwarfs of few 100 pc diameter will be only marginally resolved, unless lensed as in the object of \cite{vanzella19}.

For a better chance to distinguish true GCPs from their host,  or from other high-redshift dwarfs or close multiple GCPs or multiple knots of star formation, we will have to take advantage of the higher spatial resolution provided by lensing (such as in \citealt{vanzella19}) or of the next generation of extremely large telescopes (ELT) assisted by advanced adaptive optics. For example, the 39m European ELT will provide a $\sim 6$ times better spatial resolution compared to JWST, corresponding to $\sim 30$ pc at these redshifts.

On the side of the mass budget factor, the mere stellar mass loss via stellar winds  would account for a factor $\sim 1.7$ in GC mass reduction from formation to the present. On top of this, star evaporation and stripping via disk shocking and tidal interactions would further reduce the GC masses which according to the $N$-body simulations of \cite{webb15} could be as high as a factor of $\sim 10$ with an average of a factor $\sim 4.5$. Thus, a mass budget factor of order of 10 does not appear to be unconceivable. Again, direct counts will be the only 
way to estimate this critical factor.

With each of its pointings, NIRCam will sample a comoving volume between $z=3$ and 10 of over 160,000 Mpc$^3$. Brightest cluster  galaxies (BCG) as massive as M87, likely with a similar share of $\sim 10,000$ GCs, come with a space density of $\sim 10^{-5}$ Mpc$^{-3}$ \citep{bernardi13}, hence there is a fair chance that each NIRCam pointing will include one BCG precursor along with the precursor of the galaxy cluster hosting it. If Nature has been benign enough to make bright GCPs, we will have the opportunity to learn much about the star formation and its clustering preceding the appearance of massive galaxies, with clustered GCPs working as signposts of incipient massive galaxy formation. 

\begin{figure}
\includegraphics[width=84mm ]{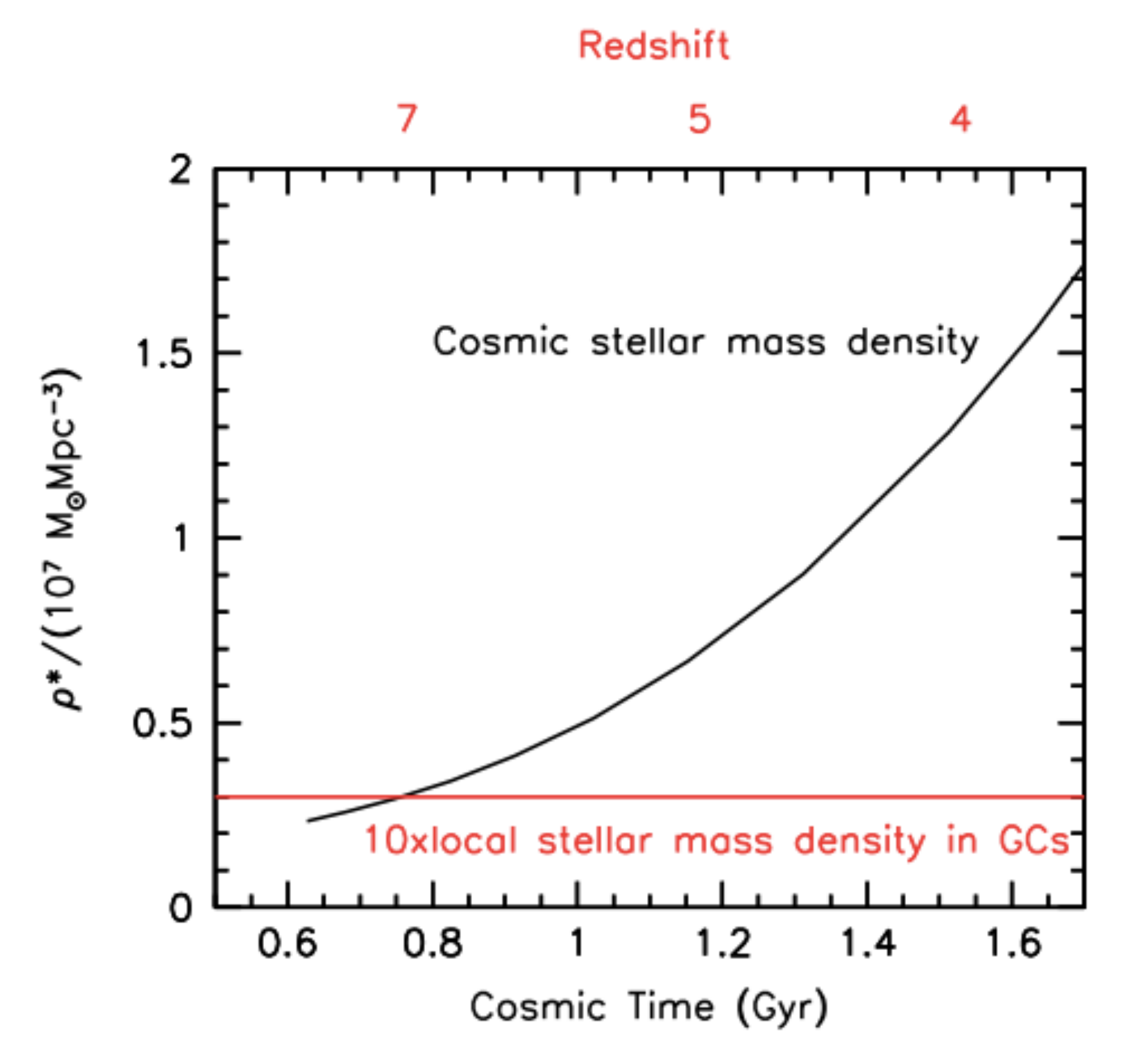}
\caption{The early evolution of the comoving cosmic star mass density as from  Madau \& Dickinson (2014) (black line), adjusted to our adopted Chabrier IMF. The red horizontal line is drawn at the level of $3\times 10^6, \msun$ Mpc$^{-3}$, corresponding to 10 times the local stellar mass density in globular clusters.
}
\label{fig:rhostargc}
\end{figure} 
  
\section{Summary  and Conclusions }
Having assumed and justified that the spectrum of young GCPs, or at least of their $\sim 50$ percent metal poor fraction, could reasonably be described by 
simple stellar population models, we have shown that:

\begin{itemize}
\item
Given the power-law shape of the spectrum of young GCPs,  both colours and fluxes/magnitudes in NIRCam passbands are fairly insensitive to the actual formation redshift. Only in the bluest passbands (namely, F70W, F90W and F115W) colours, luminosities and counts are dramatically affected by hydrogen absorption in the intervening IGM  along the line of sight.
\item
As a consequence, we show that for F150W and redder filters the expected number counts of GCPs is fairly insensitive to the actual distribution of formation redshifts, in the range $3<z<10$.
\item
Number counts depend instead critically on the actual mass distribution of GCPs, i.e., on how much more massive they were compared to their GC progeny. The ratio of the initial to present GC mass, commonly referred to as the {\it mass budget factor}, is the primary controller of the GCP number counts. For such factor being 1 (GCPs as massive as today GCs, i.e., no mass loss) NIRCam should detect of order of 10 GCPs per pointing down to mag $\simeq 30$, a number that jumps to $\sim 1,000$ if such factor is instead 10.
The recently observed number density of candidate GCPs at $z=6-8$, revealed by the HFF program, suggests that  the {\it mass budget factor} could be at least 4, i.e., GCPs being at least 4 times more massive than their local descendants, if all were to end up as GCs.
\item
For a mass budget factor of 10, GCPs should start to dominate the number counts of high-$z$ galaxies just beyond the limits currently achieved so far, i.e., $m_{\rm F160W}\simeq 29$ \citep{bouwens15}.
\item
Thus, actual number counts will set stringent constraints on the mass budget factor, providing crucial information on the formation and early evolutionary stages of GCs and helping to decipher their multiple generation phenomenon.
\item
Like some GCs today, GCPs may have been hosted by dwarf galaxies which could be detected either directly (for a recent tantalising finding see \citealt{vanzella19}) or by stacking many detected GCPs, thus characterising the environment having nursed GCs at their formation epoch.
\item
As metal poor GCs formed well before the bulk of the galaxy hosting them today, clustering of GCPs on $\sim 100$ kpc scale would mark the signpost of incipient massive galaxy formation. In this respect, reaching $\sim 1$ magnitude deeper, i.e., down to mag=31 as in a future JWST {\it Ultra Deep Field}, should boost number counts by a factor $\sim 4$, thus greatly facilitating the GCP clustering analysis.
\item
The only way of measuring GCP photometric redshifts appears to be via the drop-out technique, as bluest photons are absorbed by intervening hydrogen. Using only NIRCam, drop-outs in the filters F070W, F090W and F115W will correspond to GCPs at  $z>5$, $>6.5$ and $>8.5$, respectively. Photometric redshifts for objects at $z<5$ will have to rely on bluer data from either HST or the ground.
\end{itemize}

We also qualitatively discuss  a series of caveats concerning all these predictions, including extinction (likely affecting the metal rich fraction of GCPs),  the possibility of the IMF of GCPs being different from the assumed \cite{chabrier03} IMF, the effect of the  GCP mass being built up with multiple episodes of star formation, and finally the possible role of massive binaries. We do not explore the possibility of young GCPs hosting a supermassive star of $\sim 10^4-10^5\,\msun$, that could significantly contribute to their overall luminosity \citep{denissenkov14}.

Finally, we briefly comment on the possible role of  GCPs on cosmic reionization, an issue already mentioned in Section 1. Figure \ref{fig:rhostargc} shows the evolution of the comoving stellar mass density as a function of cosmic time as from \cite{madau14}, adjusted to our choice of the Chabrier IMF. The red horizontal line is drawn at 10 times the local mass density in GCs, which is $\sim 3\times 10^5\,\msun$ Mpc$^{-3}$. From the figure, we see that GCs, together with their possible dwarf galaxy hosts, may have dominated the cosmic mass density and, therefore, the reionization if the bulk of them formed at $z\gsim 7$, whereas their contribution would have been marginal if formed predominantly at $z\lsim 4$. This is consistent with the early estimate by \cite{ricotti02}, thus leaving open the connection with reionization, especially if the escape fraction of ionizing photons from GCPs were  close to unity as argued by \cite{ricotti02} and \cite{katz14}.

\section*{Acknowledgments}
We would like to thank Eros Vanzella, Francesco Calura, 
Enrico Vesperini and Gianni Zamorani for 
stimulating discussions and useful input on the evolution of GCs.
We also thank Rychard Bouwens for providing his data in electronic form and for useful input.
We thank the referee Massimo Ricotti for his constructive comments and suggestions.
AR, LP acknowledge support from an INAF/PRIN-SKA 2017 grant.


\appendix
\section{Evolution with time and mag-mass relations in all other  filters}

\begin{figure}
\includegraphics[width=84mm ]{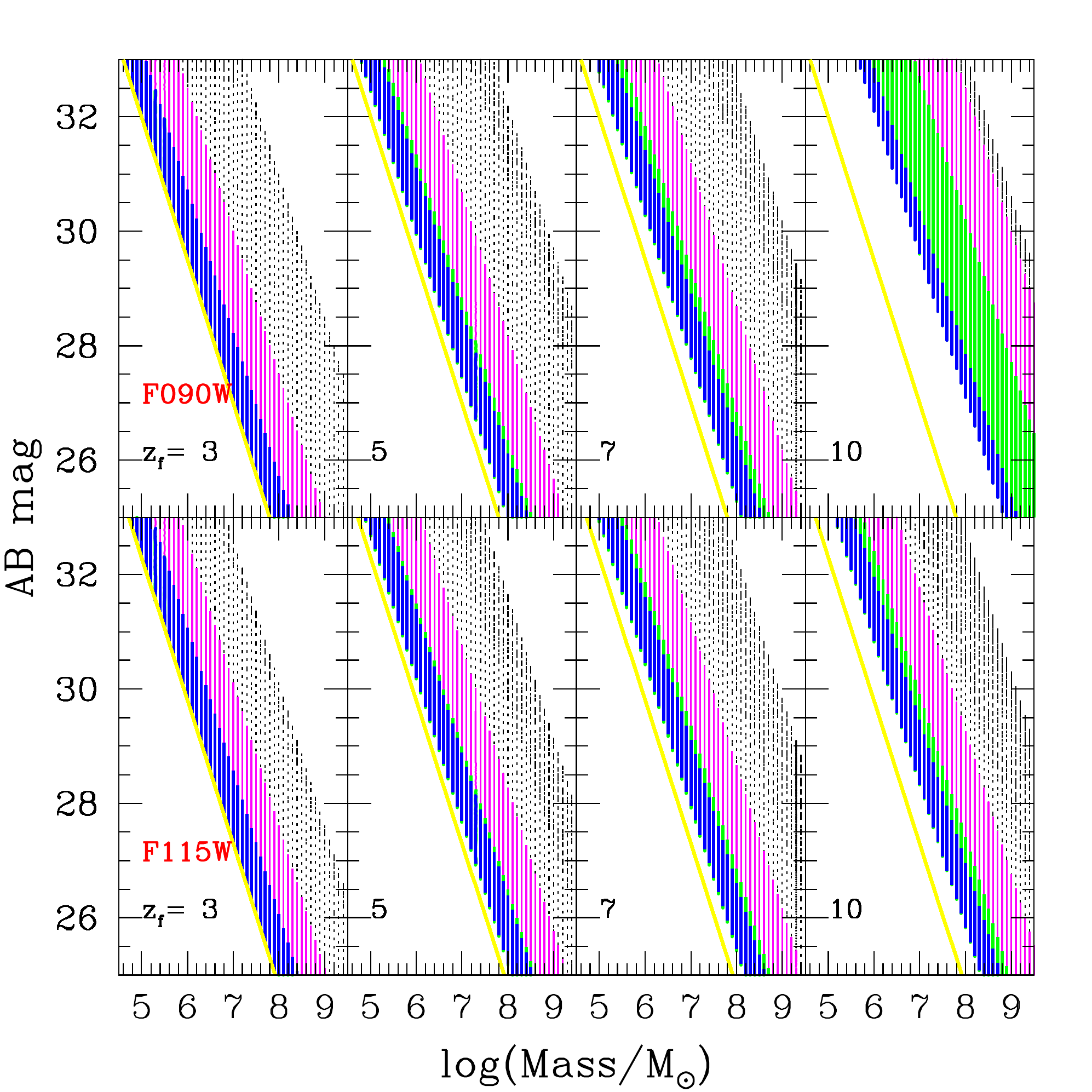}
\caption{The same as Fig. \ref{fig:mag_mass} for the  F090W and F115W bands, as indicated.}
\label{fig:mag_mass2}
\end{figure}
\begin{figure}
\includegraphics[width=84mm ]{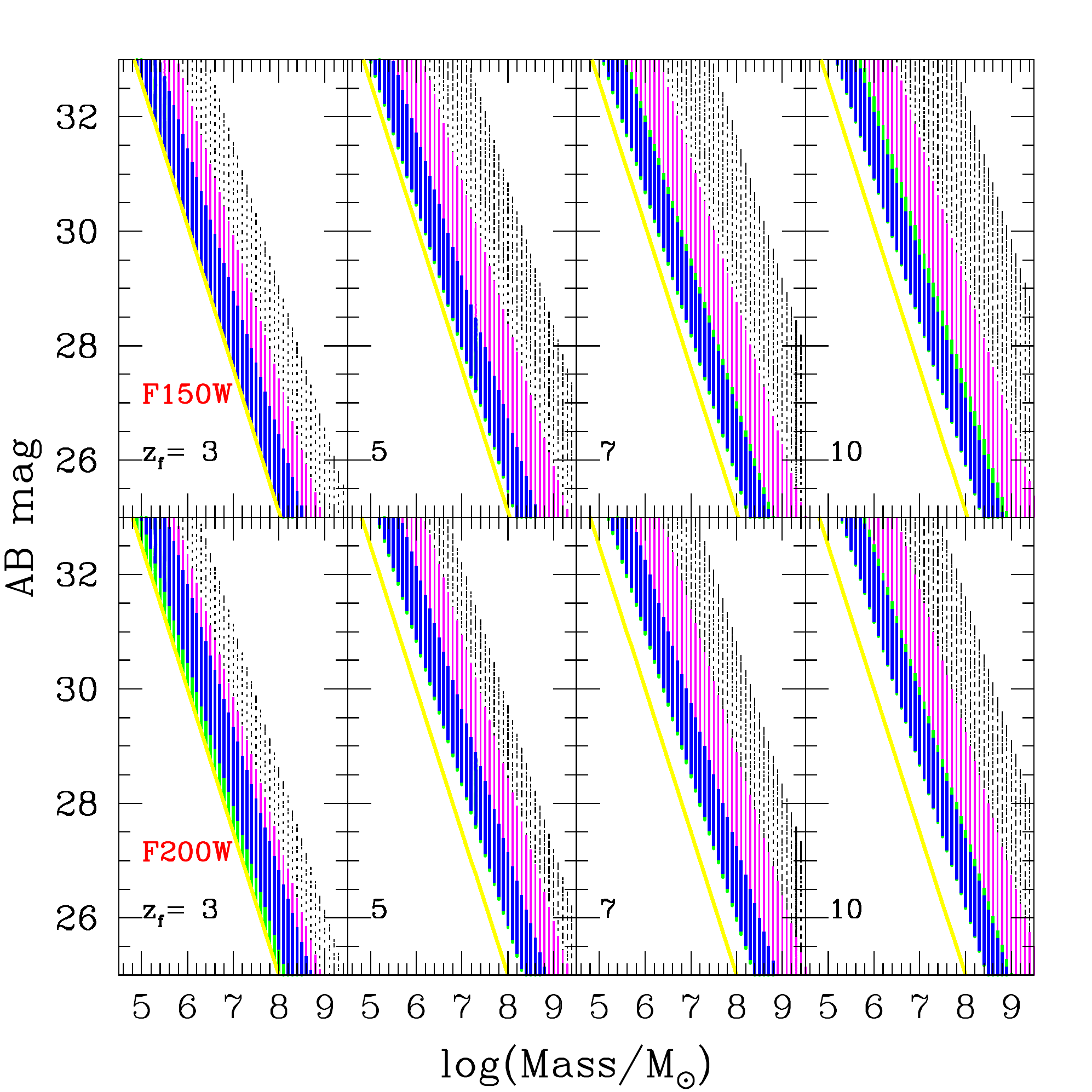}
\caption{The same as Figure \ref{fig:mag_mass} for the F150W and the F200W bands, as indicated.}
\label{fig:mag_mass3}
\end{figure}
\begin{figure}
\includegraphics[width=84mm ]{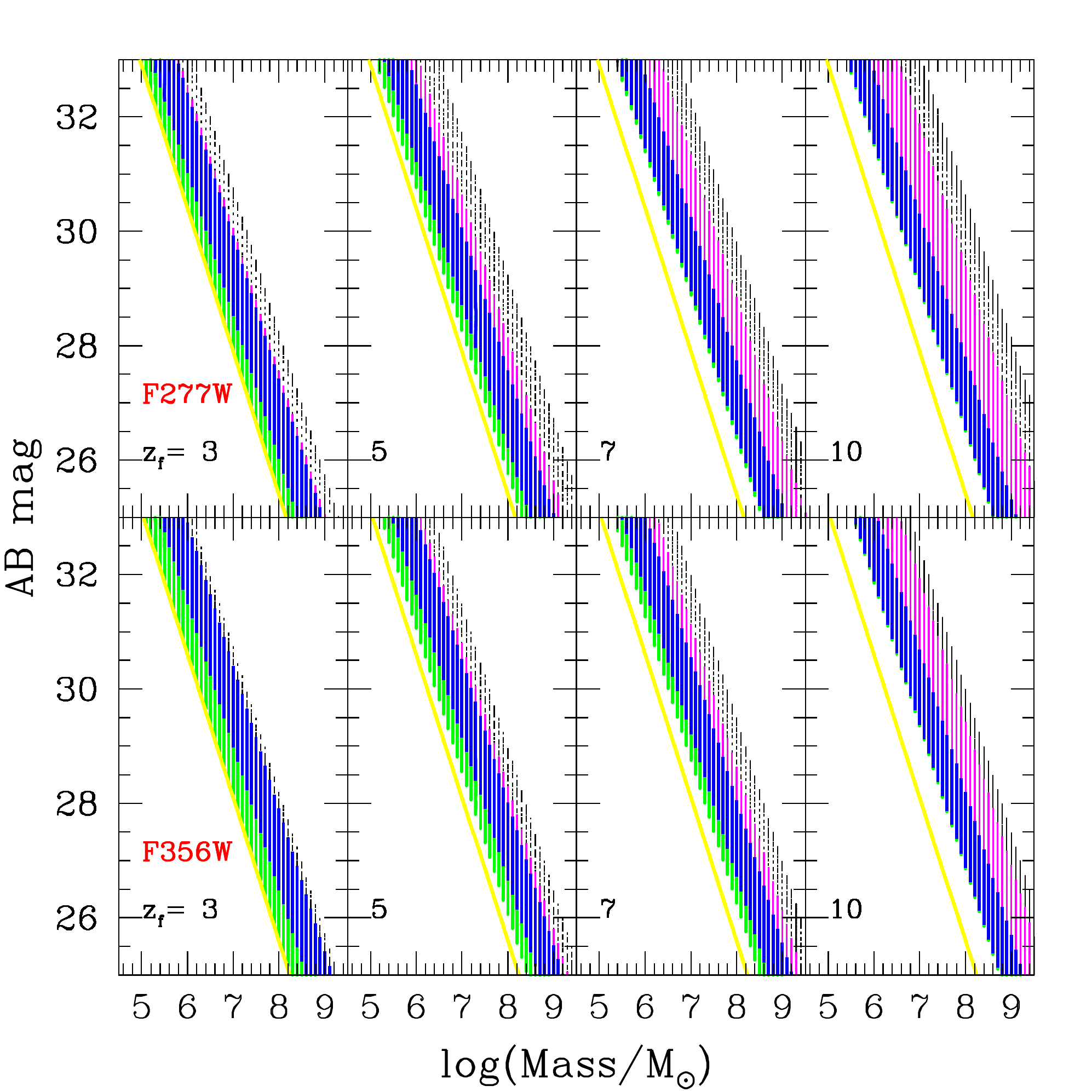}
\caption{The same as Figure \ref{fig:mag_mass} for the F277W and the F356W bands, as indicated.}
\label{fig:mag_mass4}
\end{figure}

For completeness,
Figures \ref{fig:mag_mass2}, \ref{fig:mag_mass3} and \ref{fig:mag_mass4} are analog to Figure 7, but for the other NIRCam filters.

\label{lastpage}

\end{document}